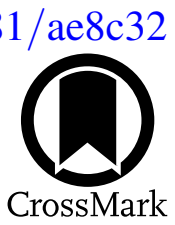

# Exploring Long-period Architectures: Four New Planet Candidates from Kepler with Periods >342 days

Matthew T. Hansen and Jason A. Dittmann
Department of Astronomy, University of Florida, Bryant Space Science Center, Stadium Road, Gainesville, FL 32611, USA; matthew.hansen@ufl.edu

## Abstract

The Kepler detection pipeline, as well as the transit method, have a bias towards shorter periods, leaving a dearth of detections at longer orbital periods. This relative lack of detections has left an incomplete picture of the architectures of exoplanet systems within the long-period regime. We have built a single transit detection pipeline, utilizing a classification convolutional neural network and the onboard spacecraft diagnostics of the Kepler spacecraft, to detect long-period planets. We apply our pipeline to all currently known planetary systems in the Kepler field hosting at least one planet with an orbital period longer than 6 days. We manually vet all new signals from our pipeline, and identify four new planetary candidates, all of which are in systems where the inner planets exhibit transit timing variations (TTVs). Two of these candidates, Kepler 1752.02 and Kepler 199.03, cause two transit events that are consistent with periods of $777.78^{+0.01}_{-0.02}$ and $505.495^{+0.004}_{-0.004}$ days, and radii of $3.55^{+0.15}_{-0.15}$ and $2.74^{+0.05}_{-0.05}R_{\oplus}$, respectively. Our remaining two candidates, Kepler 1897.02 and Kepler 1811.02, are single transit candidates with radii $4.81^{+0.20}_{-0.19}$ and $3.25^{+0.28}_{-0.30}R_{\oplus}$, respectively. The shortest orbital periods for these candidates, consistent with the Kepler dataset (gaps and coverage), are 342 days for Kepler 1897.02 and 544 days for Kepler 1811.02. The new planetary candidates, on their own, are incapable of reproducing the observed TTV signals in the inner system. Although difficult to schedule, follow-up observations are needed to further constrain the new candidates and potentially discover the planets causing the perturbations.

## 1. Introduction

After the discovery of the first exoplanets (A. Wolszczan & D. A. Frail 1992; M. Mayor & D. Queloz 1995), the field of exoplanets has seen a rapid expansion in the total number of confirmed planets, now totaling 6138 confirmed planets.[1] Currently, the most successful detection method by number of planets discovered is the transit method.

The Kepler Space Telescope, henceforth Kepler, was a space-based mission that operated from 2009–2013. Its primary science goal was to measure the frequency of Earth-like planets around Sun-like stars, $\eta$-Earth (W. J. Borucki et al. 2010). Kepler observed approximately 150,000 stars during its primary mission, allowing for a nearly 4 yr window of continuous photometric data for these stars. In 2013, a second reaction wheel ceased operating, marking the end of the original Kepler mission. However, observational data were taken across the ecliptic as the K2 mission until 2018 (S. B. Howell et al. 2014).

Another transit-method-based mission is the Transiting Exoplanet Survey Satellite (TESS). TESS is an all-sky survey that was launched in 2018 (G. R. Ricker et al. 2015). The primary mission of TESS lasted until 2020, and it is now in its extended mission. TESS was designed to detect sub-Neptune-sized exoplanets around the brightest stars to enable follow-up observations, including spectroscopy of the host stars and potentially the atmospheres of the planets. TESS observed a patch of sky in 30 day increments, called sectors. After each sector, TESS would rotate to another patch of sky. This allowed for nearly all-sky coverage, covering more stars at the expense of a lack of a longer baseline of observation for a smaller subset of stars (such as in Kepler). The TESS mission also observed the Kepler field in various sectors, providing additional data and photometric follow-up for the Kepler stars.

Along with photometric data for every target star, Kepler also provided onboard spacecraft diagnostics, made available through the mission's ancillary engineering files, throughout its 4 yr original mission. There are a total of 39 additional attributes included in these onboard spacecraft data. These attributes include the state of the reaction wheels, the temperatures throughout the spacecraft (e.g., the temperature on the Schmidt Corrector), the mean and standard deviation in the attitude pointing error about the $x$-, $y$-, and $z$-axis, among others. The original intention of the engineering files was to help in detrending stellar lightcurves by removing systematic sources of error. This was later abandoned, as cotrending basis vectors (CBVs) were found to be a better tool in correcting Kepler lightcurves. In our previous work (M. T. Hansen & J. A. Dittmann 2024), we explored machine learning (ML) techniques that combined the engineering files with photometric data and showed that there is information within these files that can boost transit classification accuracy relative to a flux-only classification scheme, and in this work, we will expand on this previous study.

The 4 yr of observed data of Kepler allows for the discovery of planets ranging from extremely short orbital periods (<1 day) to long-orbital periods (>100 days) (E. A. Petigura et al. 2013; D. Foreman–Mackey et al. 2016; S. E. Thompson et al. 2018;

[1] https://exoplanetarchive.ipac.caltech.edu/docs/counts_detail.html

K. Wang et al. 2024). A longer baseline generally also allows for detections of smaller radius planets at short orbital periods since there are more transits to boost the planet's signal-to-noise ratio (SNR) within the lightcurve. At long-orbital periods, this advantage largely disappears, as there are fewer transits available to phase-fold and boost the SNR. Since there are fewer transits, this results in a general trend of the detected planets at long-orbital periods having a larger $R_p/R_*$ (where $R_p$ and $R_*$ are the radii of the planet and host star, respectively) and, hence, a larger SNR (S. E. Thompson et al. 2018; D. C. Hsu et al. 2018).

The value of $\eta$-Earth has been sought after since before the end of the primary mission of Kepler (J. Catanzarite & M. Shao 2011) . The range of values for $\eta$-Earth has spanned over an order of magnitude in recent years, from $0.015^{+0.011}_{-0.007}$ (S. Bryson et al. 2020) at the low end to $0.58^{+0.73}_{-0.33}$ (S. Bryson et al. 2021) at the high end, indicating measurements are likely dominated by systematic corrections. Studies have applied various techniques in order to reach a value of $\eta$-Earth, such as assuming a complete sample (J. Catanzarite & M. Shao 2011), performing an injection recovery analysis on an independent pipeline (E. A. Petigura et al. 2013), an injection recovery analysis on the Kepler pipeline, and a systematic analysis on parameter uncertainties (C. J. Burke et al. 2015). Studies have attempted solutions for extending occurrence rate power laws into the Earth-like regime. The decrease in detections of Earth-sized planets at long-orbital periods has resulted in a general lack of agreement on $\eta$-Earth from previous studies, since they have required extrapolation to accommodate the falloff of detections (W. M. Farr et al. 2014; D. Foreman–Mackey et al. 2014; A. Silburt et al. 2015; W. A. Traub 2015; D. Garrett et al. 2018; G. D. Mulders et al. 2018; J. K. Zink & B. M. S. Hansen 2019; S. Bryson et al. 2020).

While understanding systematic effects in the Kepler data is important, the most straightforward approach to reducing the uncertainty in the current measurements of $\eta$-Earth is to detect more planets at longer periods. Previous studies, when detecting long-period planets, have visually inspected the lightcurves, resulting in planets with few high-SNR transits within Kepler (J. Wang et al. 2013, 2015; S. Uehara et al. 2016).

Visually inspecting lightcurves, although tedious, has some advantages over automated pipelines, such as not necessarily needing to detrend the lightcurves. Most automated detrending techniques are designed to remove long-term trends while preserving short-term features, such as transits. If a planet's period is sufficiently long, its transit duration can surpass the frequency-threshold detrending techniques used to detrend the lightcurve, resulting in the removal of the transit signal (A. Vanderburg & J. A. Johnson 2014). For example, J. Wang et al. (2015) detected multiple planets that were missed from previous studies, since automated detrending techniques also removed the transits along with the stellar variability.

Previous studies have developed techniques to systematically detect planets with long periods within the long-baseline of Kepler. D. Foreman–Mackey et al. (2016) computed a likelihood minimization of a top hat along a lightcurve, looking for single transit events. However, they required an event to be more than 25 times deeper than the background noise to be classified as a candidate in their first step, resulting in a systematic bias toward larger planets. Nevertheless, there remains a dearth of detections of long-orbital period planets.

ML has been applied to detect new exoplanet candidates (C. J. Shallue & A. Vanderburg 2018; M. T. Hansen & J. A. Dittmann 2024). ML has also been used to help validate exoplanet candidates, which has resulted in a significant speed-up over the manual inspection for validation (S. D. McCauliff et al. 2015; C. J. Shallue & A. Vanderburg 2018; S. E. Thompson et al. 2018; D. J. Armstrong et al. 2021; H. Valizadegan et al. 2022). More recently, H. Valizadegan et al. (2025) also incorporated the ancillary engineering data files of TESS, the momentum dumps, into their deep learning transit signal classification framework, `ExoMiner++`, to aid in the validation of exoplanets.

In our previous work, M. T. Hansen & J. A. Dittmann (2024), we created a single transit detection pipeline that combined the onboard spacecraft diagnostics of Kepler along with the photometry to detect new exoplanet signals. Our primary goal of building a single transit detection pipeline is to remove the dependence on phase-folding and extend the threshold of detection efficiency toward longer periods. Our pipeline utilized convolutional neural networks (CNNs) to detect locations within a stellar lightcurve that contain transits. We found that the engineering files contained information that ameliorated the pipeline's ability to detect transits within the lightcurve. We found a new long-period planet candidate that only contained a single visible transit within Kepler, representing our first steps in increasing the number of detections in the long-period regime.

Within the totality of the known exoplanet systems, a subset of them contains planets that experience transit timing variations (TTVs). TTVs are deviations from the predicted transit times of one planet from gravitational perturbations away from Keplerian motion from a third body within the system. Numerous studies have calculated and cataloged the TTV signals in Kepler (E. B. Ford et al. 2012; T. Mazeh et al. 2013; T. Holczer et al. 2016; M. Kane et al. 2019). TTVs can encode information about the planets' masses and eccentricities (E. Agol et al. 2005; M. J. Holman & N. W. Murray 2005). One can also infer the existence of another planet within a system, even if there is only one transiting planet that exhibits TTVs (E. Agol et al. 2005; M. J. Holman & N. W. Murray 2005; S. Ballard et al. 2011). As most perturbers will be at a farther distance away from the host star, compared to the already discovered planet, the use of TTVs can be crucial in detecting these planets. Previous studies have detected nontransiting planets in such a manner already (S. Ballard et al. 2011; D. Nesvorný et al. 2012, 2013).

In this work, we build upon our original pipeline, first introduced in M. T. Hansen & J. A. Dittmann (2024), and apply our newly adjusted pipeline to all systems that have precise transit timings and a period longer than twice our pipeline input. We report the identification of four new candidates within these systems and perform a series of tests to place constraints on their physical and orbital parameters.

In Section 2 we detail our dataset. In Section 3 we describe our pipeline, and our method of vetting planet candidates. In Section 4 we describe the four new candidates' parameters and list potential future transit times for possible follow-up observations. In Section 5 we discuss the possibilities of an undetected perturber and the impact of future missions. We conclude in Section 6.

## 2. Data

### 2.1. Neural Network Training Set

The NASA exoplanet archive[2] hosts all of the known confirmed and candidate planets as well as the known false positives that have been discovered in the Kepler dataset. We utilize the confirmed and candidate to train our network, and we obtain the dataset from the NASA Exoplanet Archive (NASA Exoplanet Archive 2026).[3] We use the same strategy for obtaining a training, validation, and testing set as we used in the first iteration of our pipeline, presented in M. T. Hansen & J. A. Dittmann (2024): we impose two cuts on the exoplanet catalog to create our training set: (1) an orbital period longer than 12.5 days; and (2) a transit depth less than 350 parts per million (ppm). The period cutoff ensures that no one transit appears in multiple training segments. A transit depth of 350 ppm corresponds to a 2 $R_\oplus$ planet transiting a 1 $R_\odot$ star, ensuring our training sample is representing small planet signals in lightcurves and is not dominated by easier-to-detect large planet signals. Our final training, validation, and testing set consists of 860 planets in 768 unique systems. For our training, validation, and testing sets, we impose an 80/10/10 split based on the Kepler Input Catalog (KIC) number, to ensure there is no data leakage of a single star between sets of data. This results in a total of 616 systems in our training set, 78 systems in our validation set, and 74 systems in our testing set.

To obtain the lightcurves for each star, we utilize `lightkurve,Lightkurve`. We stitch each available long-cadence quarter for a star, then normalize, and remove all *nan* values. The data that we store for each star from `lightkurve` are the time stamps, flux values, flux errors, and cadence numbers. The flux values are normalized but not detrended, so there may be stellar variability and signals at other frequencies that can create false positives for our network. We detrend each star using a best-fit spline, using the b2 statistic (A. Vanderburg & J. A. Johnson 2014). We do not mask out the transits when detrending, treating each star within the training set identical to how we would treat stars when searching blindly for new potential transit signals. The "ancillary data," also known as the engineering files, store the onboard spacecraft diagnostics throughout the lifetime of the mission. The times tamps used for the engineering files are sampled at a different frequency compared to the lightcurve's time stamps. However, each time stamp within the engineering files states which cadence it is associated with. For each photometric cadence, we take the mean for each engineering attribute reported as taking place during the associated cadence.

Our network is a binary classifier trained to predict if a section of a lightcurve provided as input contains a transit or does not contain a transit. To create our training dataset for the in-transit classification, we center a 128 cadence segment on each individual transit for the planets within our training set. We use the period and time of the first transit within Kepler for each candidate as listed on the NASA Exoplanet Archive. We also ensure there are no large data gaps within the segment by checking the time stamp difference between the first and last index of the segment. If the difference is larger than 10% of the 128 long-cadence input window, or roughly 2.93 days, we discard this segment and construct a lightcurve statement for the next transit. If our lightcurve segment passes this test, we collect the time stamps, detrended flux values, flux errors, cadence numbers, and all 39 engineering attributes associated with each cadence number. The final sizes of our in-transit data segments for training, validation, and testing are 26,534, 3341, and 3192, respectively.

To create the no-transit dataset, we first flag all time stamps within the lightcurve associated with a transit. As before, we use the MAST-provided value for the central transit time of the first transit, the period, and the duration of a transit. Then, we start on the first time stamp of the respective lightcurve and place a 128 long-cadence window. If any indices within this window are associated with a transit, we discard the window and move to the next index. If the window is free of in-transit indices, we further check to ensure there are no large data gaps present (10% of the window size). Some stars will not have any no-transit data segments (for example, if they have a short-period transiting planet as well) and, hence, contribute no learning for the no-transit classification. After a window segment has passed these checks, we collect the same data as the in-transit segments and move the window to the next set of 128 cadences. The final sizes of our no-transit data segments for training, validation, and testing are 239,879, 32,062, and 28,827, respectively.

There is an imbalance between our in-transit and no-transit data; we have 9X more no-transit segments compared to the in-transit segments. To combat this issue, we randomly downselect the no-transit dataset to have a 50/50 split in labels within the training, validation, and testing sets with the in-transit segments. The final sizes of our training, validation, and testing sets with a 50/50 split in labels are 53,068, 6682, and 6384, respectively.

Each segment undergoes data standardization before it enters into our network for training, validation, and testing. For the flux segments, we perform the same standardization as presented in A. Teachey & D. Kipping (2021):

$$\boldsymbol{F}_{\rm stand} = \frac{\boldsymbol{F} - \min(\boldsymbol{F})}{\tilde{F} - \min(\boldsymbol{F})} - 1, \quad (1)$$

where $\boldsymbol{F}_{\rm stand}$ is the standardized flux to be inputted into the network, $\boldsymbol{F}$ is the array of fluxes, $\tilde{F}$ is the median value of the flux input, and $\min(\boldsymbol{F})$ is the minimum of the flux array. The engineering files undergo a different standardization. For each engineering attribute, we take the difference between each individual point and the median over the entire 128 cadence set, and divide by the median absolute deviation (MAD) of the attribute. This equates to turning each point into a distance from the median of the attribute in units of the MAD.

### 2.2. Construction of Network

We introduced the first iteration of our pipeline in M. T. Hansen & J. A. Dittmann (2024). In that work, we built an ensemble of CNNs to identify single transit events within Kepler, utilizing photometric and onboard spacecraft data. Here, we present the second iteration of our pipeline to address some shortcomings we identified in our original version.

First, we decreased the input into our pipeline from 500 long-cadence points to 128 long-cadence points. We continue to use CNNs and fully connected neural networks (FCNNs) to take flux and onboard spacecraft diagnostics as input to make our transit classification. We first create our flux-only network, which is used as a baseline for comparison when adding the engineering features to the network. We utilize a 1D CNN structure to extract features out of the flux input, followed by an FCNN classification block to predict transits. The final neuron has a sigmoid activation function, which makes the output a value between 0 and 1 and represents a confidence level in the networks classification. A value close to 0 shows strong confidence in the no-transit classification, where a value close to 1 is strong confidence in a transit classification. We used Bayesian Optimization and Hyperband (BOHB) to tune the parameters within our network (S. Falkner et al. 2018). BOHB iterated over the number of convolutional blocks, where each block had two convolutional layers followed by a pooling layer; the number of filters and kernel size for each convolutional block; average or max pooling along with their pool size and stride length; the dropout rate; the number of neurons in the first dense layer; and the number of dense layers. For the dense layers, we halved the number of neurons from the previous layer, and BOHB decided the optimal number of dense layers with this halving requirement. Each network was trained for a maximum of 50 epochs. We used an early stopping condition that if the validation accuracy did not improve for five epochs, training was stopped, and the highest validation accuracy model was saved. The majority of networks were finished training before epoch 20. The final architecture for our flux-only network is shown in Figure 1.

[2] https://exoplanetarchive.ipac.caltech.edu/cgi-bin/TblView/nph-tblView?app=ExoTbls&config=cumulative

[3] Accessed on 2024 August 21 at 07:39, returning 4725 rows.

When adding the engineering attributes into the network, we used an FCNN block to extract features out of the nine engineering attributes. The nine attributes (as labeled in the ancillary engineering files) are: ADATTERRMX, ADATTERRMY, ADATTERRMZ, ADATTERRDX, ADATTERRDY, ADATTERRDZ, ADRW1SPD, ADRW2SPD, ADRW3SPD, and ADRW4SPD. Instead of having a separate CNN column for each engineering attribute, as in M. T. Hansen & J. A. Dittmann (2024), we combine all nine engineering attributes as input into a single FCNN block. The engineering FCNN column consisted of FCNN blocks, where each block was two dense layers of the same number of neurons followed by a dropout layer, and each subsequent block's number of neurons was half of the previous block. We used BOHB to determine the optimal number of engineering FCNN blocks, and the starting number of neurons. After the flux input passes through the CNN block, and the engineering attributes pass through the FCNN, we concatenate the outputs and input them into an FCNN to make a classification of transit or no-transit. This concatenation occurs after the dropout layer, as seen in Figure 1. Each star within our dataset, described in Section 2.3, will be passed through both of these networks for classifications of new signals.

Once the networks' trainings were completed, we calculated a receiver operating characteristic (ROC) curve to determine the optimal threshold for our classification on the testing set. We also calculate the area under the curve (AUC) to represent the overall accuracy of our models. An ROC curve varies the threshold to make a positive classification, and computes a true-positive and false-positive rate for that threshold. We choose the threshold based on which threshold places the model closest to the ideal location. For the flux-engineering model, the threshold for classification is 0.4414, and the overall model had an AUC value of 0.8. The threshold for classification, based on the ROC curve, for the flux-only model is 0.5195, with an AUC value of 0.81. The ROC curves for the two networks are shown in Figure 2.

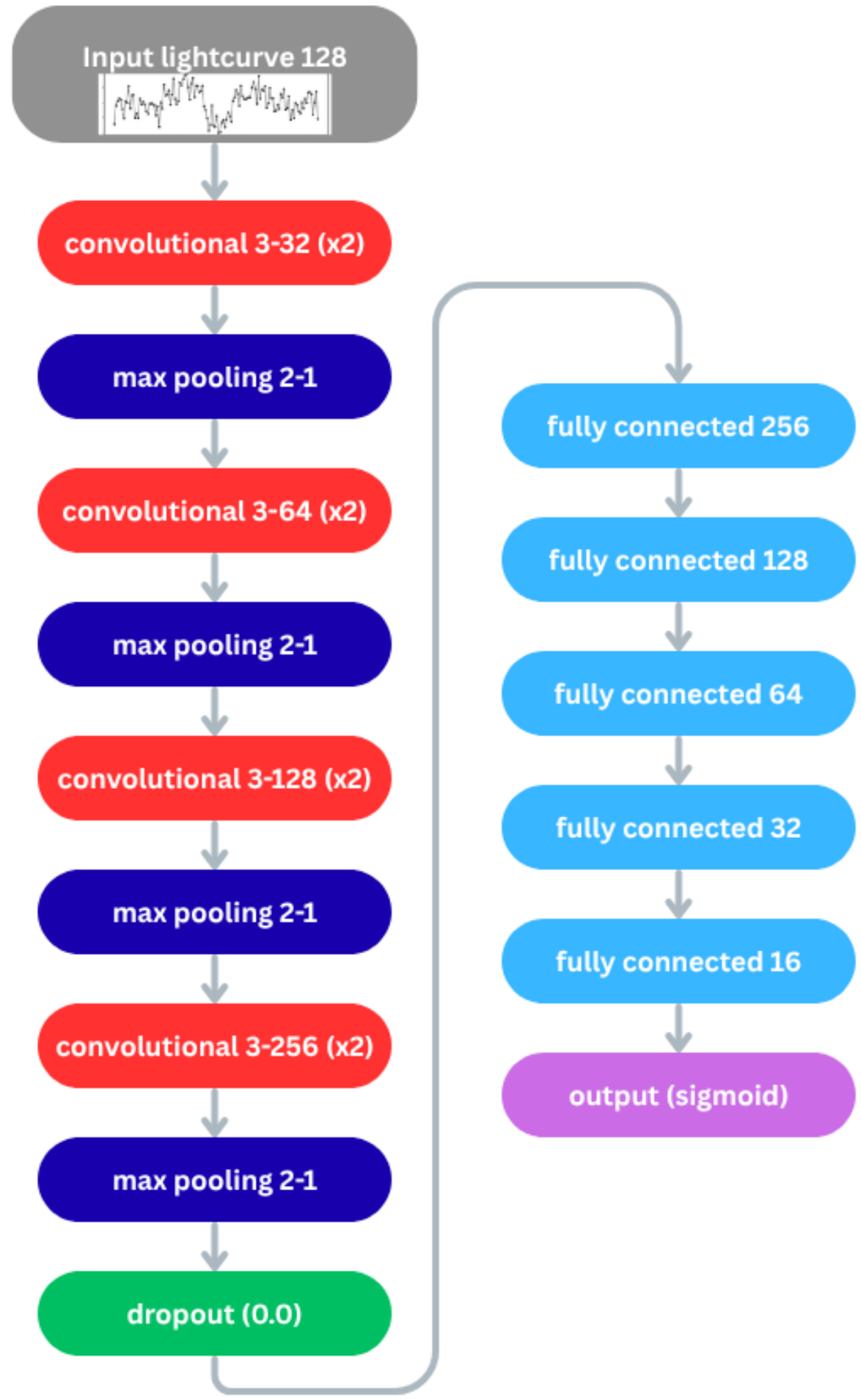

**Figure 1.** The architecture of the flux-only network with a 128 long-cadence input from the host star. The convolutional layers (red) are stated with their kernel size (number of layers; number of filters).The pooling layers (dark blue) are described with pool size (stride length). The dropout layer (green) is shown with the dropout rate, and the fully connected layers (light blue) are shown with the number of neurons.

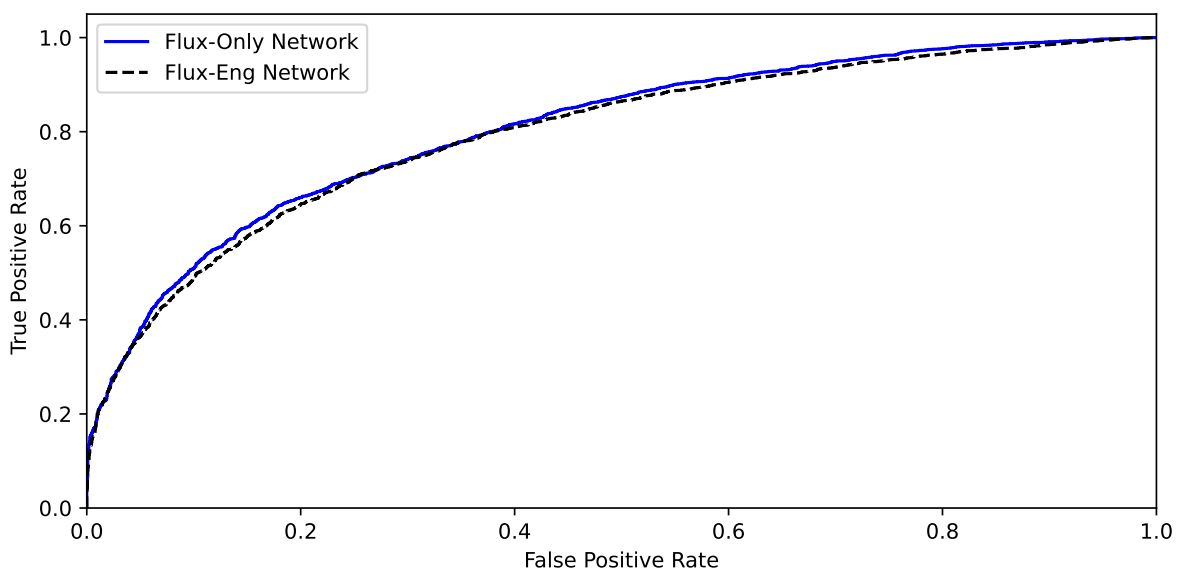

**Figure 2.** The receiver operating characteristic (ROC) curve for our flux-only (blue) and flux-engineering (black dashed) networks. With true-positive rate on the $y$-axis and false-positive rate on the $x$-axis, the ideal threshold would place the model on the $y$-axis with a true-positive rate of 100% and a false-positive rate of 0%, and an area under the curve (AUC) value of 1. The AUC values are 0.81 and 0.80 for the flux-only and the flux-engineering networks, respectively. There is a marginal difference when comparing the two models.

Both networks achieve similar results, and no significant difference is seen between them. This is in contrast to our earlier work in M. T. Hansen & J. A. Dittmann (2024), where we saw an increase in performance when using the flux-engineering network structure. This could suggest that spacecraft systematics operate on timescales longer than the chosen 128 long cadence. More research is needed to further test why a smaller input size diminishes the increase in performance in the flux-engineering structure. Since there is no increase in performance from one network to another, we use both models separately throughout this paper. This will serve as a check for any new signals if they are picked up by both models. As will be explained in Sections 3.3 and 4, we arrive at the same four new candidate long-period transiting planets with each network.

### 2.3. Dataset

The next step for our pipeline is to try to detect new, undiscovered planets. Planet candidates have a lower false-positive probability when they are within a multiplanetary system as opposed to being the only candidate orbiting the star (J. J. Lissauer et al. 2012; J. F. Rowe et al. 2014). Previous studies have shown that multiplanet systems tend to have a low mutual inclination (J. J. Lissauer et al. 2011; D. C. Fabrycky et al. 2014). For systems that are known to host at least one transiting planet, if there are other nondetected planets within their system, they have an increased probabilty of being within a transiting configuration due to their low mutual inclination. Therefore, we choose to first apply our pipeline to systems with known transiting planets to potentially uncover additional planet candidates and to explore long-period exoplanet architectures.

We use the T. Holczer et al. (2016) catalog of transit times for Kepler objects of interest (KOIs) for planetary parameters. T. Holczer et al. (2016) discarded 2091 KOIs of the 4690 KOIs presented on the archive at the time of their publication, based on certain criteria such as a low-SNR transit or a transit depth greater than 10% correlating with an increased probability of being an eclipsing binary. They analyzed a total of 2599 KOI lightcurves to obtain a catalog of all transit times for each KOI. We remove any systems with a planet with an orbital period shorter than 6 days, due to our input size into our network being 128 long-cadence points; otherwise, our pipeline would not be able to distinguish between transits of the known planets versus new signals (see Section 3 for more details). After this period cutoff, we are left with 1961 KOIs in 1605 systems. These 1605 systems represent the subsample within Kepler most likely to contain more, undetected, transiting planets since their orbital plane is already known to be transiting and known to host planets. For our sample of 1605 stars, 2% have a stellar radius larger than 3.5 $R_\oplus$. The majority of our samples' stellar radii measurements, 1498, come from the reported values in the T. A. Berger et al. (2026) catalog. If a star was not present in the T. A. Berger et al. (2026) catalog, we used the stellar radius from the NASA Exoplanet Archive data we described in Section 2.1. We note that for a very small minority of stars that some parameters are internally inconsistent with what is reported. For example, one star, KOI 5873, has a reported radius of 11.723 $R_\odot$, with an orbiting planet, KOI 5873.01, with a reported depth of 772 ppm, equating to 35.52 $R_\oplus$. However, the reported planetary radius in the exoplanet archive is $109061^{+1.748e+04}_{-3.265e+04} R_\oplus$. Moreover, the TIC catalog for this star, TIC 27773672, gives a radius of 8.80 $R_\oplus$, based on GAIA DR2 data (M. Paegert et al. 2021). None of our new reported candidates are within these stellar systems that contain highly uncertain stellar radii. Here we are just noting the existence of these outliers within our dataset.

## 3. Single Transit Detection Pipeline

### 3.1. Single Star Application

Throughout this subsection, we will walk through a test case using the exoplanet host KOI 1527 (KIC 7768451). KOI 1527 is a $0.91^{+0.04}_{-0.03} R_\odot$ star with a mass of $0.96^{+0.04}_{-0.06} M_\odot$ (T. A. Berger et al. 2026). KOI 1527 hosts one known transiting planet, KOI 1527.01, which has a period of 192.66658 $\pm$ 0.00002 days, and a radius of $3.17^{+0.13}_{-0.10} R_\oplus$ (T. Holczer et al. 2016). KOI 1527.01 has seven visible transits within the Kepler dataset, and the average single transit SNR is 9.8.

Our training set consisted of segments of lightcurves either centered about a transit, or containing no transit, and the network was asked to make a classification of "transit" or "no-transit." However, when looking for new transit events, we do not know the location of the transit time a priori to center our window. To circumvent this issue, we employ a sliding window across the lightcurve to collect data. We initialize our 128 long-cadence window on the zeroth index of a single stellar lightcurve. We check if the first and last time stamps of our window are within 10% leeway or 2.93 days, to ensure no large data gaps are present. If the window segment passes the data gap check, we collect the flux and engineering data for that window, and we slide our window one cadence over to repeat the process. Therefore, a single cadence is able to enter our network a maximum of 128 times. We continue throughout all available data for that star. For KOI 1527, this method produces 57,317 window segments to be fed into our network.

Once we collect all available window segments for a single star, we run each of them through our network to classify which window segments possibly contain a transit. For each time stamp, we count the number of times it was included in a window segment that was flagged by our network as containing a transit. Therefore, each time stamp will have a corresponding value representative of the frequency with which that time stamp was within a window containing a transit-like event. We expect every star to have a noise floor with which any given time stamp within a stellar lightcurve is spuriously flagged by our network, and any time stamps above this noise floor to have an increased probability of being a true transit-like event (see Figure 3 for a similar plot).

Due to the presence of data gaps, time stamps close to a data gap may not enter into the network a large number of times. Our network may classify each window segment the transit is in as a positive identification of a transit, resulting in 100% accuracy; however, the frequency of the classification may still be below the noise floor of the star due to the limited data surrounding the transit. For this reason, we compute the fractional classification rate for each time stamp: the total number of positive classifications divided by the total number of times the time stamp entered the network as part of a lightcurve segment. We also require a time stamp to enter the network a minimum of 80 times (out of a maximum of 128) in order to prevent statistical variation from creating false positives.

The new metric of the fractional classification rate for each time stamp allows us to label sections of the stellar lightcurve based on our network's classification. We run

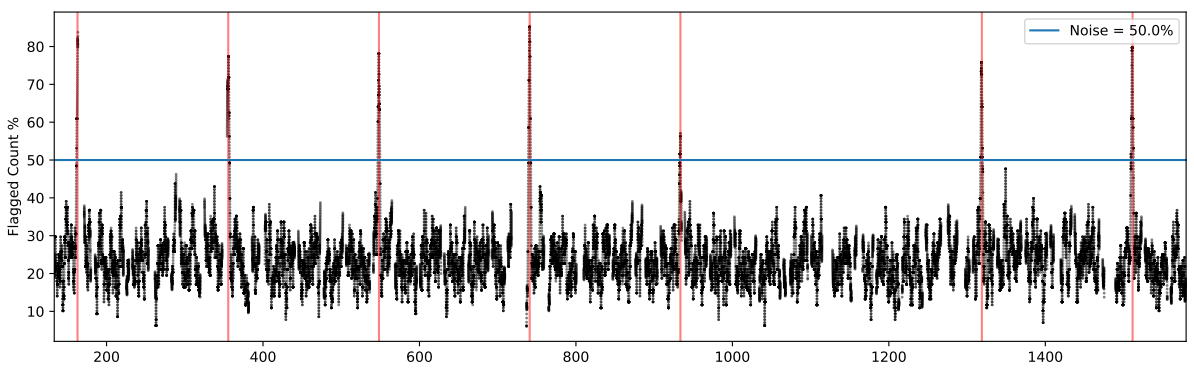

**Figure 3.** The percentage a time stamp is flagged as being within a window containing a transit compared to how often the time stamp enters the network for KOI 1527. KOI 1527.01 is a $3.17^{+0.13}_{-0.10}R_{\oplus}$ planet with a period of 192.66658 ± 0.00002 days, around a $0.91^{+0.04}_{-0.03}R_{\odot}$ star. The horizontal blue line represents the 50% threshold we use to classify significant peaks to declare a transit event. The highlighted red regions are the locations of KOI 1527.01 transits that are visible within the lightcurve. There are seven visible transits within the lightcurve, and our pipeline recovers all seven of them. One transit of KOI 1527.01 falls within a data gap and is, therefore, not highlighted. Our pipeline also correctly recovers the known period, reporting a period of 191.68 days.

`scipysignal`'s peak-finding algorithm to determine the peaks within the array (P. Virtanen et al. 2020). We define a peak in this array as any value greater than 50%, representing that our network classifies the window as containing a transit 50% of the time when said time stamp is within the window. We impose a second condition that peaks must be more than half the window size apart, or 64 long-cadence points in separation. This is to ensure the peak-finding algorithm does not double-label the same peak. Moving forward, we call any peak above this 50% threshold (and before we do any more analysis on it) a threshold crossing event (TCE) of our pipeline. In Figure 3 we show the percent frequency plot for KOI 1527. The highlighted regions are where there is a visible transit of KOI 1527.01 within Kepler. The horizontal blue line represents the 50% threshold cutoff to define a peak. Our pipeline correctly labels all visible transits of KOI 1527.01 as TCEs.

Once the peak-finding algorithm labels all peaks within the lightcurve, we can deduce an orbital period for the TCE. If there is only a single TCE, we report no period. If there is more than one TCE, we modify the percent flagged plot by casting all time stamps above the 50% threshold cutoff to 1, and all time stamps below to 0 (effectively creating boxcar pulses). We then run `astropy`'s Lomb–Scargle periodogram on the modified plot to produce the period. This ensures all signals below the threshold are suppressed, and the periodogram can only pick up signals from the TCEs. We recover a period of 191.68 days from the periodogram for KOI 1527.01. This is in strong agreement with the known period of 192.67 days. If our pipeline were to recover no peaks above the 50% threshold, we run the periodogram on the unmodified percent frequency array, and not on the modified array described above. Therefore, even though no transits were predicted above the 50% threshold, the periodogram could still identify power at their period in aggregate. As an example, this can be seen in Section 4.4 when no transits from Kepler 1897b were above the 50% threshold, but the periodogram grabbed the 1:2.5 alias period of Kepler 1897b.

### 3.2. Dataset Application

As described in Section 2.3, our dataset consists of 1961 KOIs in 1605 systems. Each star is treated in the same manner as described in Section 3.1. After each star in our dataset is fed through the flux-only and flux-engineering networks, we run the peak-finding algorithm to search for any peaks above 50% to be counted as a TCE. However, since we are looking for new signals, we must account for the known planets within the system, which may cause their own TCE. Using the transit times for the planets, as reported in T. Holczer et al. (2016), we ignore any TCE that is within half a window size (64 long-cadence points or 1.3 days) of a transit time of any known planet in the same system. We note that this method of attributing TCEs to known systems will also drop simultaneous transits, as well as transits close in time from another planet. However, these alignments are unlikely to be significant in our analysis, due to the total length of the Kepler dataset and the chance that a randomly placed new transit is in close proximity to a known transit.

For the flux-only network, after all of the TCEs that were attributed to known planets within the systems are removed, we are left with a total of 3615 TCEs within 953 systems. The flux-only network did not find any TCEs attributed to an unknown signal in 652 systems. For the flux-engineering network, after all of the TCEs that were attributed to known planets within the systems are removed, we are left with a total of 3504 TCEs within 905 systems. The flux-engineering network did not find any new TCEs within 700 systems, out of the 1605 original systems.

### 3.3. TCE Rejection

A large portion of these new TCEs originates in a small subset of stars. These erroneous TCEs are due to poor detrending of the stellar lightcurves due to the high frequency of their stellar variability. Therefore, we opt to discard any star that has more than 30 TCEs in its lightcurve. This filter removes nine stars and their corresponding 667 TCEs from the flux-engineering network's output, and seven stars with 556 TCEs from the flux-only network. Of the seven stars removed from the flux-only network, six of them overlapped with the stars removed from the flux-engineering network.

After we remove these poorly detrended stars from our TCEs, we still have 3059 TCEs in 946 systems for the flux-only network and 2837 TCEs in 896 systems for the flux-engineering network. There are known time intervals within the Kepler data that contain a common false positive among many stars simultaneously from known systematics. One of our main hopes for including the engineering files in our analysis was that the network could learn to distinguish between true transits and these types of systematics to decrease the number of reported false positives. However, some known systematics, such as a rolling band artifact, are still classified as transits in our trained neural network. We observe time stamps in which multiple stars produce TCEs at the same time and, therefore, are unlikely to be real and a systematic within the data. We computed a histogram of all peaks from our networks to inspect any time intervals with a significant increase in the amount of TCEs across all stars compared to other time bins. We used a bin size of 2.9 days, and counted all of the TCEs that fell within each discrete bin throughout the entirety of the Kepler dataset. The median number of TCEs for all bins is five TCEs per bin. If a single bin had more than five TCEs, we plotted each signal within the bin and visually checked if they were self-similar, i.e., the durations, depths in ppm, and shape were within 10% of each other for each TCE. If the TCEs were deemed as self-similar across each individual star that produced a TCE at that time stamp, we classified that time bin as containing a dataset-wide systematic. All potential

**Table 1**
The Time Bins We Declare as Systematic False Positives within the Kepler Dataset

| Bin Start, Bin End (BKJD) |
|---|
| 183.53, 186.44 |
| 198.08, 206.80 |
| 241.71, 250.44 |
| 369.72, 375.53 |
| 532.63, 535.54 |
| 631.54, 637.36 |
| 819.22, 822.13 |
| 1003.91, 1006.81 |
| 1091.18, 1097.00 |
| 1175.54, 1181.36 |
| 1184.27, 1195.91 |
| 1230.82, 1236.64 |
| 1277.36, 1280.27 |
| 1306.45, 1309.36 |
| 1338.46, 1341.36 |
| 1559.55, 1562.46 |

**Note.** Each of these bins contains a dimming event resulting in a TCE from our network. Although significant in depth, the TCEs do not resemble a transit-like shape, and the TCEs within a time bin are similar across stars. We label any TCE from our network falling within any of these time bins as a false positive.

planet signals that fall within these bins are discarded as likely false positives. We show the list of these systematic bins in Table 1. If neighboring systematic bins were determined to contain the same systematic, which may be the case since our network identifies a TCE with a $\pm 3$ day uncertainty and the discrete bin size is 2.9 days, we combined the bins into one larger bin. After discarding TCEs that fall within these bins, we are left with 811 KIC stars with 2349 TCEs for our flux-engineering network, and 847 KIC stars with 2494 TCEs for our flux-only network. In Figure 4 we show an example of one of these common time stamp systematic false positives, within the interval [1091.17941021, 1096.997665] BKJD. BKJD is defined as the Barycentric Kepler Julian Date, which is the Barycentric Julian Date (BJD) with an offset of 2454833, corresponding to 2009 January 1, the year Kepler was launched. In this figure, we show the lightcurves of six KICs, out of 120 that showed a similar systematic false positive in this time frame.

After the TCEs have passed these two checks, we manually vet each remaining TCE. We inspect both the raw and detrended lightcurves within a 6 day window centered on the TCE. From inspecting the raw lightcurve with the best-fit spline, and the resulting detrended lightcurve, we can visualize whether the TCE arose from an ill-fitted spline resulting in a false-positive identification by our neural networks. There are also TCEs that can easily be ruled out through visual inspection due to characteristic traits of a false positive, such as a sudden pixel sensitivity dropout (SPSD) event. An SPSD is typically caused by a cosmic-ray hit on the CCD and results in a sudden change in flux value in the lightcurve; see, for example, Figure 1 in D. Foreman–Mackey et al. (2016) for a visual of an SPSD.

A single TCE for a star would pass our visual inspection if it had a depth greater than 500 ppm. If a star had multiple TCEs, the visual threshold for the depth decreased, as long as the TCEs were self-similar and regularly spaced in period. For example, if one star had three TCEs all 500 days apart, with the same duration (within one or two long-cadence) and depth (within about 100 ppm), we do not require a depth greater than 500 ppm, since we can boost our confidence by observing three events. There was a total of 17 KICs that passed our visual inspection for at least one of their TCEs. These 17 KICs contained a total of 75 TCEs within our flux-engineering network and 87 TCEs within our flux-only network. However, not all of these TCEs passed our visual inspection; only 32 TCEs passed our visual inspection. Every TCE that passed visual inspection was passed in both the flux-only and the flux-engineering networks. A total of 76% of our discarded signals were signals that showed inverted behavior (brightening rather than dimming) or did not meet our depth threshold.

Once a TCE has passed visual inspection, we ensure there is not an abundance of poor quality flags within the data within/surrounding the TCE. Along with the photometric data associated with each cadence, the Kepler team also provided quality flags for each cadence to report any abnormalities that occurred during the exposure. One example of a quality flag is the rolling band artifact, which is a shift in the background of the data and masquerades as a true transit-like TCE (J. D. Twicken et al. 2018; J. Martínez-Palomera et al. 2023). This false-positive TCE arising from the rolling band artifact cannot be removed, but is only denoted through a quality flag. Therefore, for each TCE within a KIC that has passed all visual inspections, we plot the raw lightcurve surrounding the TCE and show any cadences where there was a nonzero quality flag. A quality flag of zero is a nominal cadence with no abnormalities and represents clean data. If we observed an abundance of nonzero quality flags surrounding the TCE, we discarded said TCE as a false positive. After the quality flag check, five KICs remained with seven TCEs from the flux-engineering network and eight TCEs from the flux network.

We now ensure that the total data for a KIC is consistent with the parameters that its TCE(s) predicts. For example, if a single KIC has two TCEs that are self-similar (i.e., same duration within one to two cadences, and depth within 100 ppm), and 100 days apart, we should see more transit-like events linearly spaced from the observed TCEs by 100 days. We perform this check since our pipeline may have missed some portion of real transit signals within the lightcurve that are consistent with the TCEs that have been found with our system. One KIC, out of the five that have passed all previous tests, failed this test. For this object, we observed two TCEs that were self-similar at 1334 BKJD and 1491 BKJD. However, we observe no other hints of a transit feature within the lightcurve 157 days (the separation in time between the two TCEs) away from each TCE, nor do we find any transit-like event at any alias of the period separation (aside from transits that come from the known planets within the system).

The four KICs that passed all of the above tests are KIC 12406749 with TCEs at 746.55, and 1524.33 BKJD; KIC 5113822 with TCEs at 453.14 and 1464.13 BKJD; KIC 4067925 with a TCE at 1250.11 BKJD; and KIC 5005121 with a TCE at 554.44 BKJD. Figures 5, 6, 7, and 8 show the Kepler lightcurves corresponding to the TCEs that passed our visual inspection, respectively. Overlaid in each plot is the corresponding best-fit transit model, which is explained in Section 4.1.

The next step is to investigate whether any of these signals can be explained as astrophysical false positives. We build upon the methodology of S. E. Thompson et al. (2018), where they checked for ephemeris matching and centroid offsets, and we add checks for blending from a background star, along with a search within TESS to confirm periods of the known planets and attempt to locate more transits from the new candidates to confine their periods. We follow J. L. Coughlin et al. (2014) for determining if a new signal matches an existing known signal. Using the catalog of J. L. Coughlin (2021), we compute a matching criterion between the time of the newly found transit, and period (if we detect more than one transit) among all Kepler KOIs, Kepler eclipsing binaries (EBs), ground-based (EBs), and TCEs. Our matching criteria are the same as in J. L. Coughlin et al. (2014):

$$\Delta P = \frac{P_A - P_B}{P_A}, \qquad \Delta T = \frac{T_A - T_B}{P_A}$$
$$\Delta P' = \mathrm{abs}(\Delta P - \mathrm{int}(\Delta P)), \quad \Delta T' = \mathrm{abs}(\Delta T - \mathrm{int}(\Delta T))$$
$$\sigma_P = \sqrt{2}*\mathrm{erfcinv}(\Delta P'), \quad \sigma_T = \sqrt{2}*\mathrm{erfcinv}(\Delta T') \tag{2}$$

where erfcinv() is the inverse complementary error function, and subscripts $A$ and $B$ represent objects A and B (where, in our case, object A would be the new signal, and B would be every other signal in the catalog of J. L. Coughlin 2021). To flag the new signal as a match, $\sigma_P$ and $\sigma_T$ have to be greater than 3.5, and 2.0, respectively. These values were found by J. L. Coughlin et al. (2014), empirically. These values also depend on which object has a larger value for $P$ and $T$; therefore, we also check for any matches where we treat the catalog as object A and our new signal as object B, the opposite of our original check. All KICs within our system passed this check. KIC 5113822 had three other KICs appear within the ephemeris matching; however, the transit events within the other KICs were days away from the observed TCEs in KIC 5113822. Moreover, they do not match the profile of the TCE as observed in KIC 5113822.

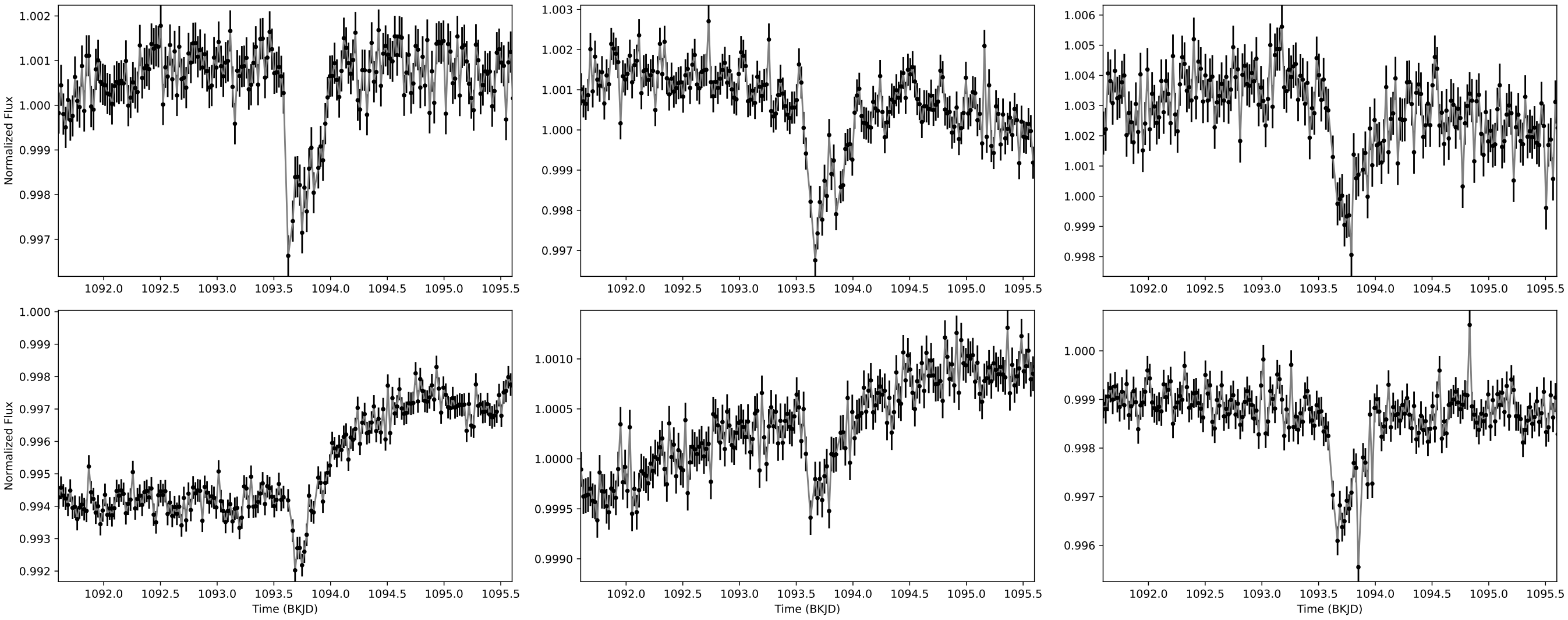

**Figure 4.** Six cases of an example systematic false positive, within the time interval [1091.18, 1097.00] BKJD. We discovered 120 total stellar lightcurves with a similar dimming event within this time interval, suggesting a systematic within the data rather than an astrophysical origin. The six KICs shown here are: KIC 10797460, KIC 12644822, KIC 6435936, KIC 6599919, KIC 8313667, and KIC 6960913. Additionally, many of the signals, while significant in depth, do not resemble a transit shape with a clear ingress, baseline, and egress. We classify the time interval [1091.18, 1097.00] BKJD as a systematic bin. We label all peaks flagged by our pipeline within a systematic bin as a false positive. Table 1 gives the complete list of all time intervals we deemed as a systematic false-positive bin.

A centroid offset would be indicative of a background eclipsing binary that was not properly detrended and is masquerading as a transit around the target star. To perform our centroid offset false-positive check, we utilize the software `vetting` to give the likelihood of a significant centroid offset resulting in the TCEs (C. Hedges 2021). The software computes the centroid of the pixels within the given aperture from `lightkurve`, and compares the $x$ and $y$ centroid positions in and out of transits using a t-test to test if the distributions are consistent. All four KICs passed the centroid offset test, with `vetting` giving no statistically significant offset during the time of the TCEs.

Although checking for blending from a nearby star may be incorporated/indirectly measured in others (such as the centroid offset), we manually check nearby stars as a superfluous measurement. We plot all stars within a 180″ radius to see if any stars contain similar features during the flagged TCE time stamps. The centroid offset test, through `vetting`, is able to give the maximum distance the blended star can be at, given the change in centroid and transit depth. However, there is no significant centroid offset for any of the KICs, so we check a constant distance of 180″. If none of the stars contain any features similar to or that could be a possible parent of the observed TCE, we claim the target signal is not caused by blending issues. All KICs passed the blending false-positive check.

When possible, we check within TESS data for observations of the target star. From the four target stars presented here, only one of them has a visual magnitude below 15, with a magnitude of 13.60. Therefore, the majority of stars are above the upper magnitude limit for the standard TESS pipeline. Nevertheless, we check the target stars for transits within TESS. The two goals for doing so are to (i) observe transits of the known planet signals to further constrain their periods, and hence their TTV signals, as well as try to (ii) observe the new planets' transits. The latter is statistically unlikely due to the short sectors of TESS, 30 days, with the long periods of the new candidates >342 days. Moreover, the sensitivity of TESS may not allow us to push to the radius regime in which we found these signals. Nevertheless, we attempted to search TESS for the known and new planets' signals, but were unable to detect any transit events.

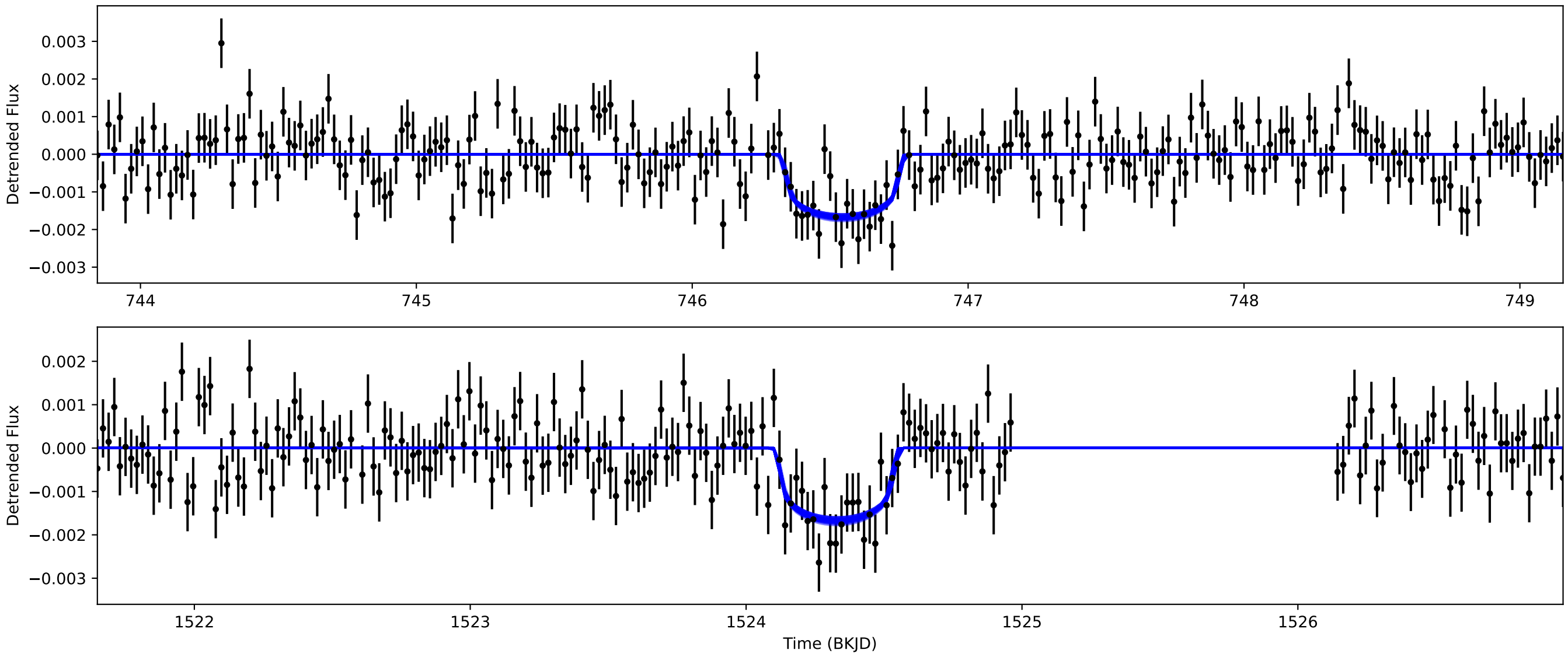

**Figure 5.** The new planetary candidate, Kepler 1752.02's, two transits at 746.55 and 1524.33 BKJD. Plotted is the detrended lightcurve, with error bars from Kepler (black), along with the transit fit produced by `batman` (blue). We show the top 400 fits to represent the uncertainty of the fit. The Markov Chain Monte Carlo (MCMC) produces a best-fit radius of $3.55^{+0.15}_{-0.15}R_{\oplus}$ for Kepler 1752.02, assuming a stellar radius of 0.788 $R_{\odot}$.

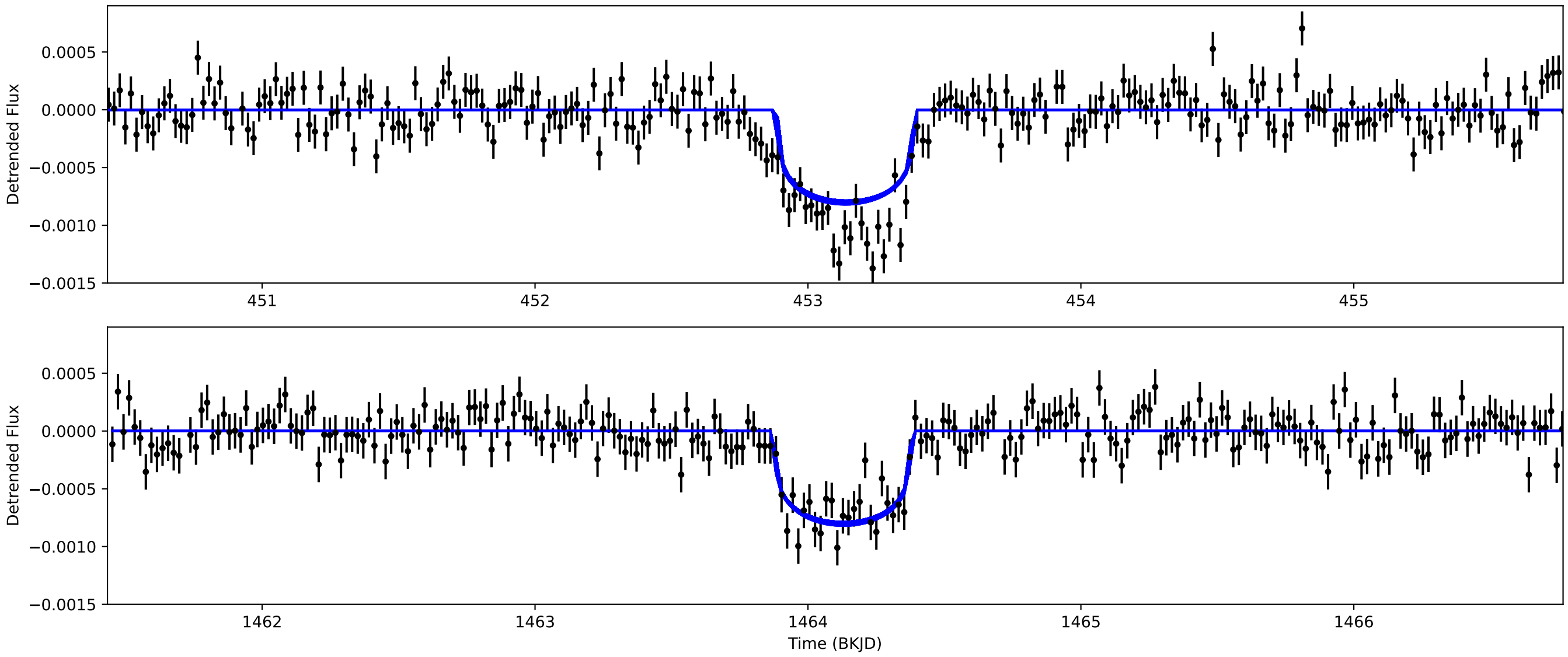

**Figure 6.** The new planetary candidate, Kepler 199.03's, two transits at 453.14 and 1464.13 BKJD. Plotted is the detrended lightcurve, with error bars from Kepler (black), along with the transit fit produced by `batman` (blue). We show the top 400 fits to represent the uncertainty of the fit. The MCMC produces a best-fit radius of $2.73^{+0.07}_{-0.06}R_{\oplus}$ for Kepler 199.03, when using a stellar radius of 0.927 $R_{\odot}$.

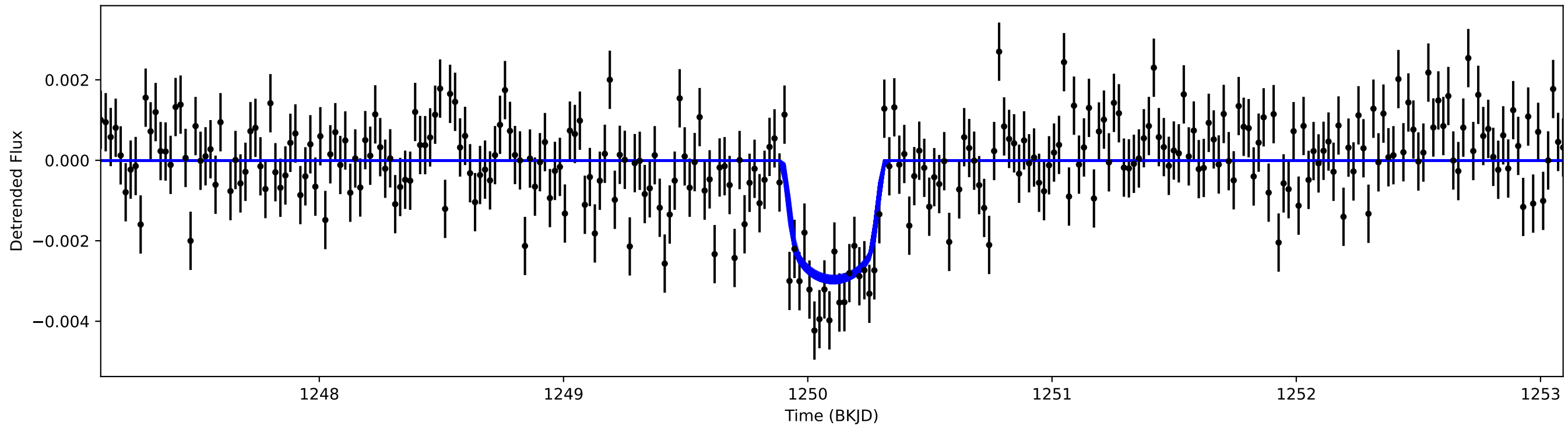

**Figure 7.** The new planetary candidate, Kepler 1897.02's, single transit at 1250.11 BKJD. Plotted is the detrended lightcurve, with error bars from Kepler (black), along with the transit fit produced by `batman` (blue). We show the top 400 fits to represent the uncertainty of the fit. The MCMC produces a best-fit radius of $4.81^{+0.20}_{-0.19}R_{\oplus}$ for Kepler 1897.02 based on a stellar radius of 0.786 $R_{\odot}$.

## 4. Discussion of New Candidates

### 4.1. Model Fitting

Three of the four systems contain only one previously known planet, and the fourth contains two previously discovered planets. We fit the known planets as well as the new candidates in our transit models.

We use the transit model producing software package, batman, to produce a transit given a set of planetary parameters (L. Kreidberg 2015). The transit software requires input for the following variables for a single planet's transit model: time of inferior conjunction, $T_0$; orbital period, $P$; planet to star radius ratio, $R_p/R_*$; the scaled semimajor axis, $a/R_*$; inclination, *inc*; eccentricity, *ecc*; argument of periastron, $\omega$; and limb-darkening coefficients for a specific model of stellar limb darkening.

We use a Markov Chain Monte Carlo (MCMC) to minimize our log likelihood function to obtain transit parameters, along with their uncertainties, for all planets and host stars in our sample. We utilize the MCMC package, emcee, to explore the parameter space (D. Foreman–Mackey et al. 2013). We use the transit models produced by batman to simultaneously fit all planets within the system, and use the reported errors for each flux cadence from Kepler within our log likelihood function.

The orbital period and the semimajor axis for any given planet are related through Kepler's third law, involving the stellar radius and mass. Instead of fitting over the stellar radius and mass, we fit over the stellar density since we can obtain the scaled semimajor axis as a function of period and stellar density. This reduces the overall number of free parameters in our MCMC, allowing for faster convergence.

For our limb-darkening model and parameters, we use the Power-2 limb-darkening coefficients, $gK$ and $hK$, as reported in A. Claret & J. Southworth (2022). To obtain the initial guess for our MCMC, we obtain the stellar surface gravity, effective temperature, and metallicity from T. A. Berger et al. (2026) to obtain the Power-2 coefficients, as reported in A. Claret & J. Southworth (2022).

For the planetary parameters of the known planets, we use the reported values in T. Holczer et al. (2016) as our initial guess. The initial guess for the new candidate was done by eye for $R_p/R_*$, and was chosen as 90° for the inclination. We choose to hold *ecc* and $\omega$ fixed to 0 and $\pi/2$, respectively, for all planets. This ensures dynamical stability for all systems presented here. We do not have a sufficient amount of data to explore nonzero eccentricity solutions, such as through priors presented in D. R. Anderson et al. (2011) and J. Eastman et al. (2013), that couple the two by $\sqrt{e}\,\sin(\omega)$ and $\sqrt{e}\,\cos(\omega)$. However, we note that at these orbital separations, it is not expected for any natal eccentricity to be damped to zero.

T. Holczer et al. (2016) provided calculated transit times, based on a linear ephemeris, along with the observed minus calculated $(O - C)$ for each transit of each KOI. There are five known planets within the four systems we analyze here. Four of the five planets have $O - C$s spanning roughly 100 minutes in magnitude, and the fifth has a max $O - C$ of less than 25 minutes. Each planet's $O - C$ values are shown in Section 5.1. Every new candidate proposed in this paper is unable to excite the inner planetary orbits to produce the observed $O - C$s while maintaining zero eccentricity. If we were to let the eccentricities vary based on the priors in D. R. Anderson et al. (2011) and J. Eastman et al. (2013), we could produce the magnitudes of the inner planets' O-Cs. We used the $N$-body simulator REBOUND to test if the orbital configuration was stable with the nonzero eccentricity solutions (H. Rein & S.-F. Liu 2012). Inputting the mean anomaly would associate each planet with the corresponding transit times, and simulated for up to 1 x $10^9$ orbits of the inner planet, with an early stopping condition of a planet close encounter, or ejection. These REBOUND simulations show that orbital parameters consistent with the observed TTV signal produce a dynamically unstable system (i.e., within 1 x $10^9$ orbits of the inner planet, there is either a collision or an ejection of a planet). We are unable to fit the magnitude of the TTV signal while maintaining dynamical stability of the system when letting eccentricity explore nonzero parameter space. Therefore, we hold eccentricities to 0 for all planets within our models.

For our priors for each parameter in our MCMC, we use Gaussian priors on the stellar density and the periods of the known planets within the system. For all other parameters, we place uniform priors where the bounds were held to keep the system physical. This included placing bounds for $R_p/R_*$ between 0 and 1, and the inclination of the system to be a fully transiting planet and not grazing. Therefore, we place bounds of $[\arccos((1 - R_P/R_*)/(a/R_*)), 90]$ on the inclination. For the power-2 limb-darkening coefficients, we placed 10% uniform priors on the values presented in A. Claret & J. Southworth (2022).

Two of our new candidates have multiple visible transits within Kepler and, hence, a likely suite of periods can be inferred. For the initial guess of the period, we take the linear separation distance between the two TCEs. We visually inspect the lightcurves at each alias to check if there was a similar transit-like event with similar depth (within 100 ppm) and duration (within one to two cadences). We also checked if there were any large data gaps within a small window around predicted locations of transits from the alias period. If there were large data gaps, we could not guarantee our initial period prediction, as the alias period would produce the same number of observed transits. The aliases that were checked were the $1/x$, where $x$ is an integer from 2–10 (e.g., for an initial separation of TCEs of 10 days, the $1/2$ alias would correspond to a period of 5 days). We can place a strong prior on this new-found period to ensure the MCMC stays within the observed period. However, the remaining two planetary candidates only contain a single visible transit within the entirety of the Kepler dataset. We can derive a period estimate based on the transit duration; however, this is a degenerate space that would also depend on *ecc*, $\omega$, and *inc*. Although we hold *ecc* and $\omega$ fixed in our analysis, there is still the degeneracy between *inc* and $P$. Therefore, for our single transiting candidates, we place a loose bound on the new planet's period between the shortest orbital period for a circular orbit to produce the observed transit time and 1400 days, representing almost the complete duration of the Kepler dataset. The shortest orbital period was found by deriving what circular orbit period would produce the transit duration, assuming an inclination of 90°, representing the longest possible chord length for a transit.

The Kepler dataset rules out many of the periods within the loose prior, because we only observe a single transit, and many periods within this prior produce a model that would contain multiple transiting events visible in extant Kepler data. We can calculate all possible periods that are allowed within the dataset that produces only one visible transit. To do this, we iteratively check which combinations of data gaps allow for a specific period to align that still produce the only visible transit observed. Within the two subsections of single transit candidates (Section 4.5 and Section 4.4), we show the allowed periods within this wide prior.

For our MCMC analysis, we utilized 64 walkers to explore the parameter space for each system. We utilized the University of Florida's high-performance supercomputer, HiPerGator, to run the MCMC in parallel across 64 CPUs. We ran each system in 5000-link increments, analyzing progress after each run. A single link took 7–8 s to run, so each 5000-link run took roughly 10–12 hr. At the start of each new iteration, we used the last links of the previous run as the starting positions for the new run. Each system was stopped after the 5000-link increment, once we had passed 50 times the largest integrated autocorrelation time, $\tau$, for the system's parameters, as our measure of convergence for the MCMC analysis.

Although we hold eccentricities to zero for all planets within our transit model fitting and do not expect the new candidates to be the perturber under these conditions, we run a TTV fitting code to ensure our assumption. We use an MCMC with `TTVFast` to produce a final $O-C$ plot for all planets within a system for a given set of orbital parameters. `TTVFast` requires a stellar mass, and planetary mass, period, inclination, eccentricity, argument of periastron, and mean anomaly. For stellar mass, we use the values presented in T. A. Berger et al. (2026) with a Gaussian prior. For planetary mass, we use the mass–radius power law from J. F. Otegi et al. (2020) for a volatile-rich atmosphere using the best-fit radius for each planet from our transit fits. We place an absolute prior on planetary mass from a 2$\sigma$ deviation given by the mass–radius power law. We hold eccentricity to 0, once again, for all planets within the system. We hold inclinations to the best-fit value that was found in our transit modeling. We let the mean anomaly vary in hopes of finding a combination of angles that could produce the observed TTV signal.

### 4.2. Kepler 1752

KIC 12406749, or Kepler 1752, is a $0.788^{+0.033}_{-0.028}R_{\odot}$, $0.818^{+.041}_{-0.043}M_{\odot}$, $5211^{+102}_{-105}$ K star with a Kepler magnitude of 15.792 (T. A. Berger et al. 2026). Kepler 1752 hosts one known transiting planet, Kepler 1752b, with a period of $56.35876 \pm 0.00001$ days and a radius of $4.59^{+0.19}_{-0.16}R_{\oplus}$ (T. Holczer et al. 2016). Kepler 1752b has TTVs with an amplitude on the order of 50 minutes (T. Holczer et al. 2016).

There are 25 potential transits of Kepler-1752b within the Kepler dataset. Three of these potential transits fall within data gaps. T. Holczer et al. (2016) gave transit times for 23 transits of Kepler-1752b, only discarding two transits. T. Holczer et al. (2016) fit a transit model to the last few points leading into a data gap as the sixth transit of Kepler-1752b; however, we see no indication that these points are a part of a transit. Therefore, we do not fit those points for a transit, and we discard that epoch in our analysis. The effects from this transit can be seen in Figure 9, where epoch 5 is an obvious discontinuity from the $O-C$ trend from the other transits.

Our flux-only pipeline recovers all visible transits of Kepler 1752b and reports a period of 18.81 days, the 1:3 alias of the period of Kepler 1752b. Our flux-engineering pipeline recovers all but one transit of Kepler 1752b, and reports a period of 28.15 days, the 1:2 alias of the known period of Kepler 1752b. Moreover, the flux-engineering pipeline does not classify some of the systematic TCEs that the flux-only pipeline reports. Our pipeline also finds two new potential transit events in Kepler 1752's lightcurve at 746.5 and 1524.3 BKJD, shown in Figure 5. Moving forward, we refer to this candidate as Kepler 1752.02. In this paper, we differentiate the new candidates from the known inner companions by calling the new candidates Kepler XXXX.0X, and not a continuation of the Kepler XXXXb nomenclature. We observe two transits of the new candidate, and, therefore, can obtain a period for the new planet candidate of 777.7 days. We find no hints of an alias period within the data when checking integer ratios for an alias period, either from other transit-like dips within the lightcurve or alignment with data gaps.

Following the methods described in Section 4.1, we first transit-fit the planetary system using an MCMC and `batman`. The best-fit parameters for Kepler 1752b and Kepler 1752.02 are listed in Table 2. The predicted period for Kepler 1752.02 is $777.78^{+0.01}_{-0.02}$ days, which would make Kepler 1752.02 within the top 20 longest-period transiting planets discovered within Kepler, if confirmed. The circular orbit transit fit for Kepler 1752.02 is shown in Figure 5. The best-fit radius for Kepler 1752.02, using a stellar radius of 0.788 $R_{\odot}$, as in T. A. Berger et al. (2026), is $3.55^{+0.15}_{-0.15}R_{\oplus}$.

The next step, following Section 4.1, is to attempt to fit the TTV signal of Kepler 1752b. We have no mass constraints on either Kepler 1752b or Kepler 1752.02; therefore, we used the mass–radius power law presented in J. F. Otegi et al. (2020) to obtain an estimate of the masses. The initial estimates for the masses of Kepler 1752b and Kepler 1752.02 were 15.50 $\pm$ 4.99 $M_{\oplus}$ and 12.01 $\pm$ 4.06 $M_{\oplus}$. From the best-fit transit timings produced by `ttvfast`, we are able to obtain a best-fit linear ephemeris for the timings and compute the TTV signal for the `ttvfast` timings. The best fit for reproducing the TTV signal of Kepler 1752b is shown in Figure 9. As expected, the circular orbit of Kepler 1752.02 cannot cause the perturbations we observe in Kepler 1752b's transit timings.

Using the `batman` best-fit parameters for $T_0$ and $P$, along with their respective uncertainties, we calculated the future transit times of Kepler 1752.02. Table 3 lists the predicted future transit times of Kepler 1752.02. The linear ephemeris we used for Kepler 1752.02 is $746.55 + 777.78^{*}n \pm (0.01 + n^{*}0.015)$ BKJD, where $n$ is the number of transits that have occurred since the first observed transit on 746.55 BKJD.

### 4.3. Kepler 199

Kepler 199, or KIC 5113822, is a $0.927^{+0.030}_{-0.025}R_{\odot}$, $0.976^{+.040}_{-0.039}$ $M_{\odot}$, $5790^{+105}_{-111}$ K star with a Kepler magnitude of 13.5950 (T. A. Berger et al. 2026). Kepler 199 hosts two known transiting planets, Kepler 199b and 199c, with periods of $23.6422 \pm 0.0003$ days and 67.09, and radii of $3.36 \pm 0.36$ $R_{\oplus}$ and $3.58 \pm 0.43$ $R_{\oplus}$, respectively (J. F. Rowe et al. 2014). Kepler 199b has TTVs on the order of 100 minutes, and Kepler 199c has TTVs on the order of 10 minutes (T. Holczer et al. 2016), shown in Figure 10.

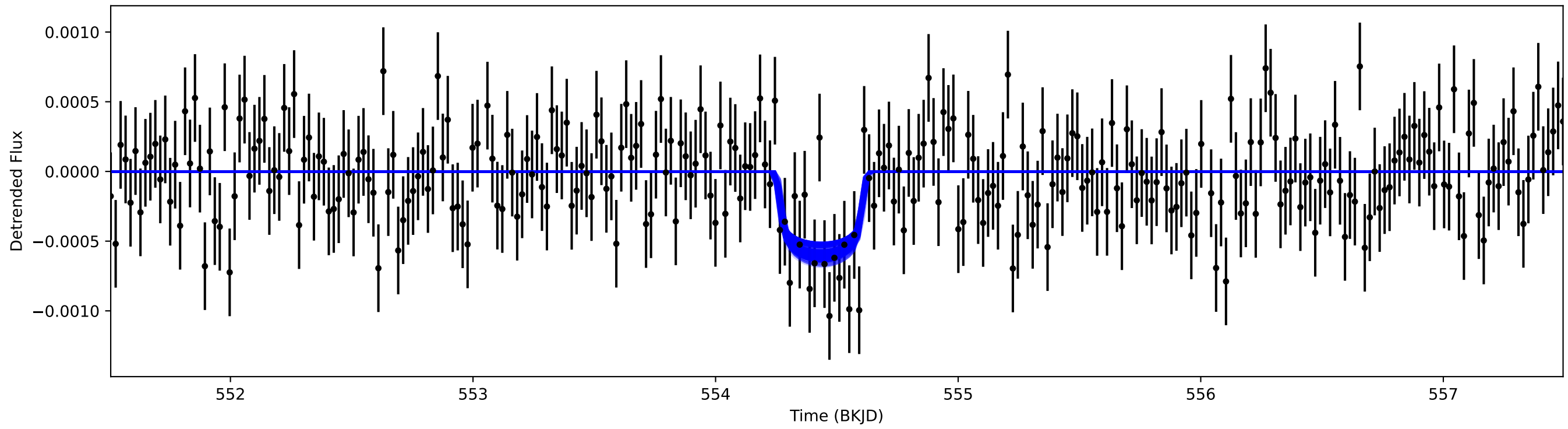

**Figure 8.** The new planetary candidate, Kepler 1811.02's, single transit at 554.44 BKJD. Plotted is the detrended lightcurve, with error bars from Kepler (black), along with the transit fit produced by batman (blue). We show the top 400 fits to represent the uncertainty of the fit. The MCMC produces a best-fit radius of $3.25^{+0.28}_{-0.30}R_{\oplus}$ for Kepler 1811.02 based on a stellar radius of 1.110 $R_{\odot}$.

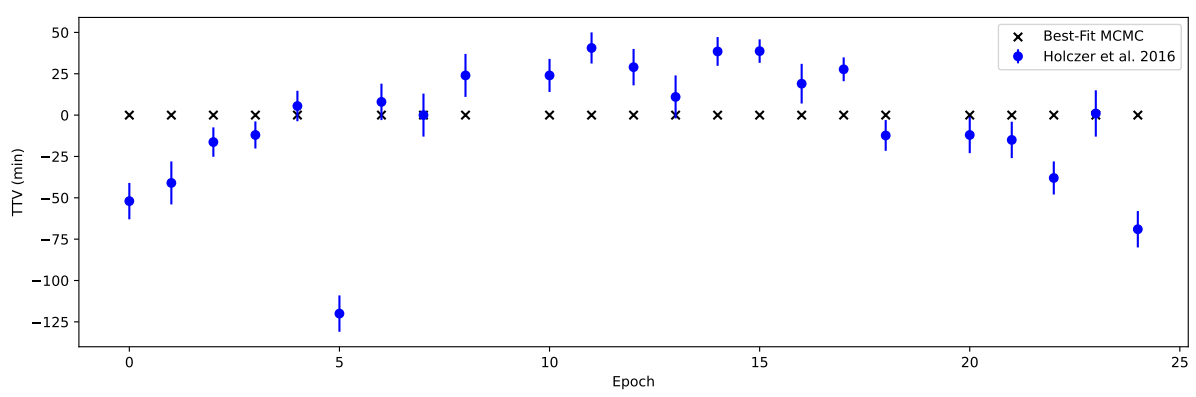

**Figure 9.** The observed TTV signal of Kepler 1752b as reported in T. Holczer et al. (2016) (blue), and the best-fit TTV signal of Kepler 1752b when using a circular model for Kepler 1752.02 (black). We remove epoch 5 from our analysis, since the linear ephemeris places the transit into a large data gap. The fifth epoch was originally assumed to be the last three data points leading into the data gap in the T. Holczer et al. (2016) fit, causing the large discrepancy in the observed TTV signal. We suggest it is more likely that these data points are not in-transit data points, as originally modeled.

**Table 2**
Best-fit Parameters for Kepler 1752, Kepler 1752b, and the New Candidate Kepler 1752.02 with an MCMC and batman

| Parameters | Kepler 1752b | Kepler 1752.02 |
|---|---|---|
| $T_0$ (BKJD) | $222.320^{+0.001}_{-0.002}$ | $746.55^{+0.01}_{-0.01}$ |
| Period (Days) | $56.358764^{+0.000008}_{-0.000008}$ | $777.78^{+0.01}_{-0.02}$ |
| Inclination (Deg) | $89.52^{+0.06}_{-0.06}$ | $89.880^{+0.006}_{-0.007}$ |
| $R_p/R_*$ | $0.0473^{+0.0008}_{-0.0008}$ | $0.041^{+0.002}_{-0.002}$ |
| $R_p$ ($R_{\oplus}$) | $4.07^{+0.07}_{-0.07}$ | $3.55^{+0.15}_{-0.15}$ |
| $\rho_*$ (g cm$^{-3}$) | $1.631^{+0.143}_{-0.142}$ | |
| $gK$ | $0.795^{+0.028}_{-0.051}$ | |
| $hK$ | $0.782^{+0.039}_{-0.056}$ | |

**Note.** We present the best-fit parameters assuming a circular orbit, $e = 0$ and $\omega = \pi/2$. Our best-fit radius for Kepler 1752b is in agreement with the presented values in T. Holczer et al. (2016). To calculate $R_p$, we use the stellar radius value 0.788 $R_{\odot}$ from T. A. Berger et al. (2026).

There are 46 and 15 transits within Kepler 199's lightcurve from Kepler 199b and Kepler 199c, respectively. Our flux-engineering and flux-only pipeline recovers all visible transits within the lightcurve for both planets. Although there are 61 total visible transits of the known planets within the lightcurve, upon visual inspection, we noticed seven transits with issues. One transit of Kepler 199b was within a data gap where the $O - C$ signal was fitted to the last two cadences entering the gap (similar to Kepler 1752b; see Section 4.2), and six other transits had increased photometric scatter (ranging from 200–500 ppm) within their transits that could alter an MCMC fitting. Therefore, we discard those seven transits when

**Table 3**
Expected Future Center Transit Times of the Next Five Future Transits of the New Planetary Candidate, Kepler 1752.02

| Kepler 1752.02 |
|---|
| 2028-01-31 07:29 ± 3:32 |
| 2030-03-19 02:16 ± 3:53 |
| 2032-05-04 21:04 ± 4:15 |
| 2034-06-21 15:52 ± 4:37 |
| 2036-08-07 10:40 ± 4:59 |

**Note.** The times are presented in GMT, in YYYY-MM-DD hh:mm ± hh:mm format. The transit duration of Kepler 1752.02 (the new signal) is 11.26 ± 0.25 hr.

performing transit model fitting. For Kepler 199b, the transits we discard are epochs 6, 14, 25, 38, and 46. For Kepler 199c, the transits we discard are epochs 5 and 9. The epochs were counted starting from 0, and based on the linear ephemeris of predicted transit times.

Along with the recovery of the known transits, our pipeline recovers two new TCEs at 453.14 and 1464.13 BKJD. The separation between the two TCEs is 1011 days. However, we observe a large data gap in the middle of the two TCEs, which suggests that a period of 505 days is also permitted by the data. We report the solution for the new candidate Kepler 199.03, assuming this 505 day orbital period.

Our best-fit transit model parameters for the Kepler 199 system from our MCMC are shown in Table 4. The reported values for Kepler 199b and 199c are within the uncertainty of the values reported by T. Holczer et al. (2016). Our new candidate, Kepler 199.03, has a best-fit radius of $2.74^{+0.05}_{-0.05}R_{\oplus}$, calculated using a stellar radius of 0.927 $R_{\odot}$. The resultant transit fit for Kepler 199.03 is shown in Figure 6.

We have no mass constraints on Kepler 199b or 199c. We are unable to TTV fit this system based on the current data and, therefore, do not have an upper limit on their mass values. Our initial guesses for the mass estimates used in the TTV fitting are from the J. F. Otegi et al. (2020) mass–radius power law. The initial mass guesses are $10.62^{+8.57}_{-5.87}$, $11.74^{+9.74}_{-6.55}$, $8.58^{+6.51}_{-4.63}M_{\oplus}$ for Kepler 199b, 199c, and 199.03, respectively. We hold eccentricity to 0 for all planets within the system, as there is insufficient data to constrain the eccentricities. As expected, the circular orbits cannot cause the observed perturbations in Kepler 199b and 199c, as seen in Figure 10.

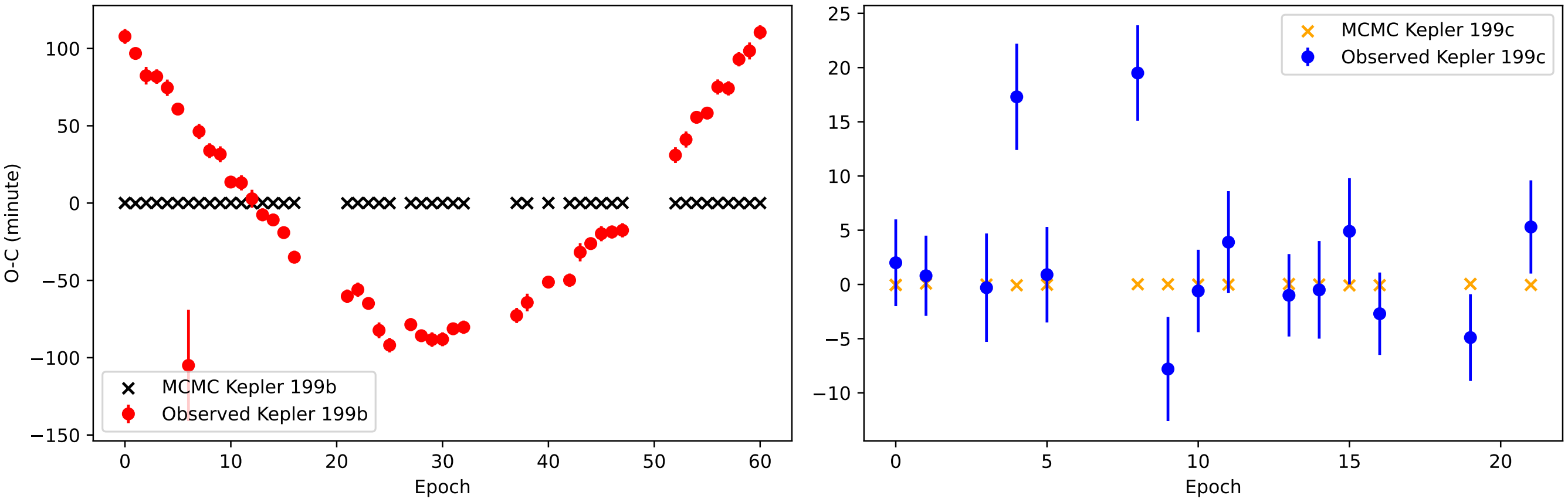

**Figure 10.** Left: the observed TTV signals of Kepler 199b, as reported in T. Holczer et al. (2016) (red), and the best-fit TTV signal of Kepler 199b (black) when using a circular model for Kepler 199b, 199c, and 199.03. Right: the observed TTV signals of Kepler 199c as reported in T. Holczer et al. (2016) (blue), and the best-fit TTV signal of Kepler 199c (orange) when using a circular model for Kepler 199b, 199c, and 199.03. We remove epochs 6, 14, 25, 38, and 46 of Kepler 199b and epochs 5 and 9 of Kepler 199c from our analysis since the transits are abnormal in their shape, depth, and/or duration and can bias our transit fits.

**Table 4**
Best-fit Parameters for Kepler 199, Kepler 199b, Kepler 199c, and the New Candidate Kepler 199.03 with an MCMC and `batman`

| Parameters | Kepler 199b | Kepler 199c | Kepler 199.03 |
|---|---|---|---|
| $T_0$ (BKJD) | $148.9491^{+0.0004}_{-0.0006}$ | $146.565^{+0.001}_{-0.002}$ | $453.137^{+0.006}_{-0.006}$ |
| Period (Days) | $23.642079^{+0.000027}_{-0.000004}$ | $67.093428^{+0.000003}_{-0.000005}$ | $505.495^{+0.004}_{-0.004}$ |
| Inclination (Deg) | $89.102^{+0.075}_{-0.027}$ | $89.515^{+0.033}_{-0.018}$ | $89.850^{+0.009}_{-0.005}$ |
| $R_p/R_*$ | $0.0311^{+0.0002}_{-0.0002}$ | $0.0331^{+0.0003}_{-0.0003}$ | $0.0271^{+0.0005}_{-0.0005}$ |
| $R_p$ ($R_\oplus$) | $3.14^{+0.02}_{-0.02}$ | $3.35^{+0.03}_{-0.03}$ | $2.74^{+0.05}_{-0.05}$ |
| $\rho_*$ (g cm$^{-3}$) | | $1.10^{+0.07}_{-0.02}$ | |
| $gK$ | | $0.80^{+0.02}_{-0.03}$ | |
| $hK$ | | $0.69^{+0.03}_{-0.04}$ | |

**Note.** We present the best-fit parameters assuming a circular orbit, $e = 0$ and $\omega = \pi/2$. Our best-fit radii for Kepler 199b and 199c are in agreement with the presented values in T. Holczer et al. (2016). To calculate $R_p$, we use the stellar radius value 0.927 $R_\odot$ from T. A. Berger et al. (2026).

We calculated the future transit times of Kepler 199.03. Table 5 lists the predicted future transit times of Kepler 199.03. The linear ephemeris we used for Kepler 199.03 is $453.137 + 505.495^{*}n \pm (0.006 + n^{*}0.004)$ BKJD, where $n$ is the number of transits that have occurred since the first observed transit on 453.137 BKJD.

### 4.4. Kepler 1897

KIC 4067925, or Kepler 1897, is a $0.786^{+0.034}_{-0.030} R_\odot$, $0.775^{+0.044}_{-0.39} M_\odot$, $4904^{+94}_{-90}$ K star with a Kepler magnitude of 15.911 (T. A. Berger et al. 2026). Kepler 1897 hosts one known transiting planet, Kepler 1897b or KOI 3066.01, with a period of 24.220470 ± 1.4 x $10^{-5}$ days and a radius of 2.44 ± $0.70 R_\oplus$. Kepler 1897b has TTVs on the order of 100 minutes (see Figure 11; T. Holczer et al. 2016).

There are 45 transits of Kepler 1897b within the Kepler dataset. Of those 45 transits, T. Holczer et al. (2016) cited 19 of them as insignificant for detection in the Kepler dataset. They performed an $\mathcal{F}$-test between the transit model for a single transit against a constant flux model, and ruled out all transits with a $p$-value larger than 0.025. Our flux-only and our flux-engineering pipelines were unable to recover any

**Table 5**
The Next Eight Predicted Transit Times of the New Candidate Kepler 199.03 with a Period of $505.495^{+0.004}_{-0.004}$ Days

| Kepler 199.03 |
|---|
| **2026-11-07 13:50 ± 1:23** |
| 2028-03-27 01:42 ± 1:29 |
| **2029-08-14 13:35 ± 1:34** |
| 2031-01-02 01:28 ± 1:40 |
| **2032-05-21 13:21 ± 1:46** |
| 2033-10-09 01:13 ± 1:51 |
| **2035-02-26 13:06 ± 1:57** |
| 2036-07-16 00:59 ± 2:03 |

**Note.** The bolded times are the predicted transit times that also correspond to the 1010 day period, and detection of the transit would not break the period degeneracy. The times are presented in GMT, in YYYY-MM-DD hh:mm ± hh:mm format. The transit duration for Kepler 199.03 is 12.75 ± 0.25 hr.

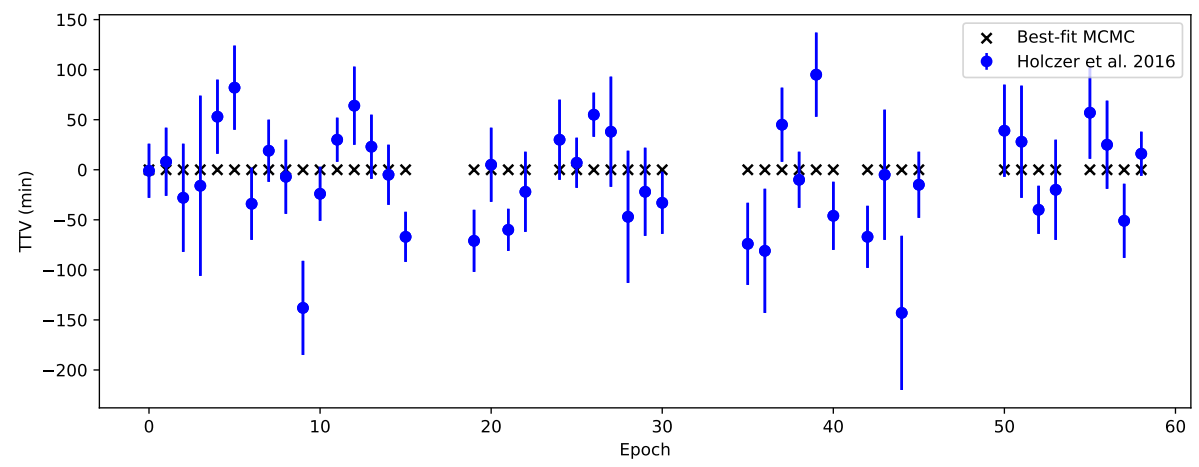

**Figure 11.** The observed TTV signal of Kepler 1897b as reported in T. Holczer et al. (2016) (blue), and the best-fit TTV signal of Kepler 1897b when using a circular model for Kepler 1897.02 produced by `TTVFast` (black). We are unable to reproduce the observed TTV signal of Kepler 1897b with the new candidate Kepler 1897.02 as the perturber with an eccentricity of 0.

individual transits from Kepler 1897b. However, our flux-engineering pipeline recovered a period of 9.7 days from the periodogram of the network's data flags, the 1:2.5 alias of the known period, suggesting that we are still able to recover this planet candidate despite no individual transit being recovered in isolation.

Our pipeline recovered one TCE within the stellar lightcurve, corresponding to a new TCE at 1250.1 BKJD, shown in Figure 7. We find no other hints of a similar TCE (i.e., a TCE is similar in shape and duration) anywhere else in the lightcurve, making this a single transit candidate. As such, we cannot place precise bounds on the new candidate's period;

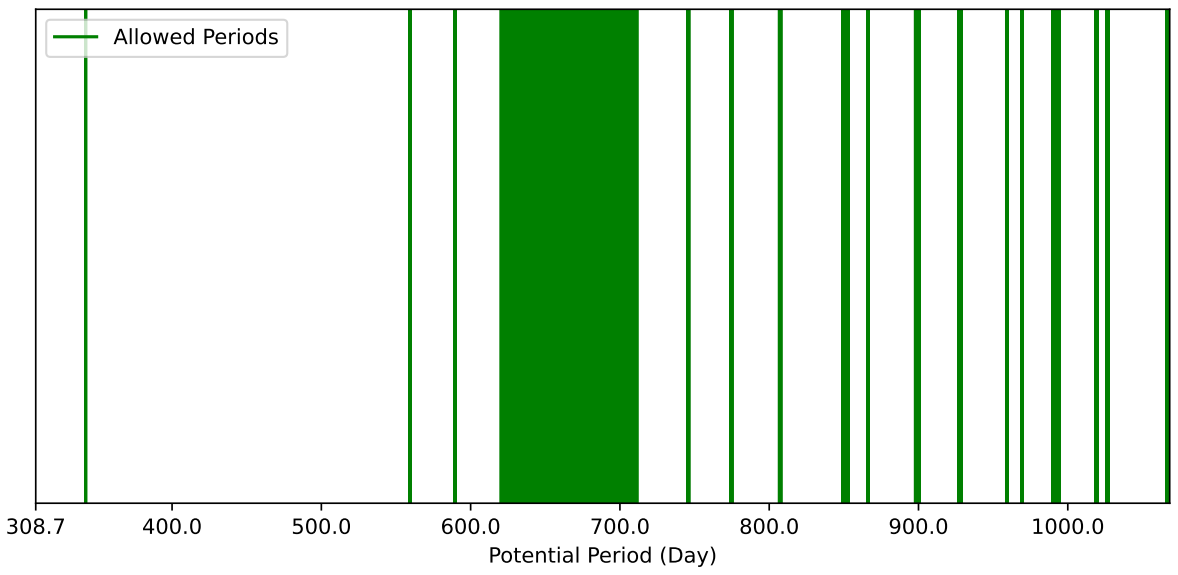

**Figure 12.** The allowed periods with the observed data gaps within Kepler for Kepler 1897.02. The minimum period, assuming a circular orbit and an impact parameter of 0 (i.e., the planet transits across the equator of the star), is 308.68 days. The green-highlighted regions are allowed periods where there are large enough data gaps to hide the other transits of Kepler 1897.02 for that specific period.

however, we can place a lower bound of 308 days for the period for a circular orbit based on the transit duration of 10.5 hr, as described in Section 4.1.

As described in Section 4.1, we can state the allowed periods within Kepler based on the data gaps present that result in only one visible transit of Kepler 1897.02. The shortest possible period, assuming a circular orbit, is 308.68 days; however, that period is not allowed with the given data, as we would expect to see other transits within the Kepler observations, but we do not observe them. The shortest allowed period, given the data gaps, is $342.136 \pm 0.001$ days, as there is a 10 hr data gap at 223.50 BKJD and large data gaps at 565 and 907 BKJD. We show the allowed periods for Kepler 1897.02 in Figure 12.

We choose the period $559.5 \pm 0.22$ days as our initial guess for Kepler 1897.02 in our transit fitting. This period corresponds to the second-shortest period allowed by data gaps. Although there is a significantly larger 90 day data gap in the time series for this target, which is correlated with the allowed period range of 610–700 days. There is a large range of periods allowed for this candidate due to this gap, which is also the statistically most likely period range. However, for convergence purposes, we have chosen the smaller period to allow our MCMC to converge within a smaller period range. We find a best-fit radius of $2.54^{+0.08}_{-0.08} R_{\oplus}$ for Kepler 1897b, within agreement of T. Holczer et al. (2016). The new candidate, Kepler 1897.02, has a best-fit radius of $4.81^{+0.20}_{-0.19} R_{\oplus}$. The resultant transit fit is shown in Figure 7, and the best-fit parameters for the entire system are presented in Table 6 when assuming a period of $559.5 \pm 0.22$ days for the initial MCMC guess.

When assuming a period of $559.42^{+0.19}_{-0.19}$ days for the new candidate, the period of Kepler 1897.02 is roughly 23 times the period of the confirmed inner planet, Kepler 1897b. Assuming an eccentricity of 0 for both planets, they are too far apart to cause the perturbations we observe for the inner planet. The MCMC fit with `TTVFast` confirms this, as there are no perturbations in Kepler 1897b's transit timings, as can be seen in Figure 11.

We calculated the future transit times of Kepler 1897.02. Table 7 lists the predicted future transit times of Kepler 1897.02 for each potential period that is allowed by the visible data gaps within Kepler. We show the next four potential central transit times for each allowed period.

**Table 6**
Best-fit Transit Parameters for Kepler 1897, Kepler 1897b, and the New Candidate Kepler 1897.02 with an MCMC and `batman`

| Parameters | Kepler 1897b | Kepler 1897.02 |
|---|---|---|
| $T_0$ (BKJD) | $170.651^{+0.001}_{-0.002}$ | $1250.105^{+0.005}_{-0.006}$ |
| Period (Days) | $24.22047^{+0.00001}_{-0.00001}$ | $559.42^{+0.19}_{-0.19}$ [One Potential Period; see Figure 12] |
| Inclination (Deg) | $88.83^{+0.07}_{-0.08}$ | $89.853^{+0.010}_{-0.010}$ |
| $R_p/R_*$ | $0.0296^{+0.0010}_{-0.0010}$ | $0.056^{+0.002}_{-0.002}$ |
| $R_p$ ($R_{\oplus}$) | $2.54^{+0.08}_{-0.08}$ | $4.81^{+0.20}_{-0.19}$ |
| $\rho_*$ (g cm$^{-3}$) | $1.64^{+0.18}_{-0.18}$ | |
| $gK$ | $0.73^{+0.05}_{-0.03}$ | |
| $hK$ | $0.81^{+0.06}_{-0.05}$ | |

**Note.** We present the best-fit parameters assuming a circular orbit, $e = 0$ and $\omega = \pi/2$. Our best-fit radius for Kepler 1897b is in agreement with the presented values in T. Holczer et al. (2016). To calculate $R_p$, we use the stellar radius value 0.786 $R_{\odot}$ from T. A. Berger et al. (2026). We report one potential period for Kepler 1897.02 that is allowed by the data gaps within Kepler (see Figure 12 for other potential periods).

### 4.5. Kepler 1811

KIC 5005121, or Kepler 1811, is a $1.110^{+0.066}_{-0.053} R_{\odot}$, $1.030^{+0.061}_{-0.78} M_{\odot}$, $5838^{+153}_{-164}$ K star with a Kepler magnitude of 15.1650 (T. A. Berger et al. 2026). Kepler 1811 hosts one known transiting planet, Kepler 181b, with a period of $76.422849 \pm 2.0$ x $10^{-5}$ days and a radius of $3.10^{+0.18}_{-0.15} R_{\oplus}$. Kepler 1811b has TTVs on the order of 100 minutes, and a synodic period of 188 days, as can be seen in Figure 13. The planet was analyzed by T. Holczer et al. (2016) and confirmed by H. Valizadegan et al. (2022) through an ML classifier.

Kepler 1811b has 19 transits within the Kepler dataset. T. Holczer et al. (2016) flagged one transit, epoch 13, as not being significant in its detection with their transit model. Our flux-only pipeline recovered seven transits of Kepler 1811b, and a period of 38.18 days, the 1:2 alias period of Kepler 1811b. The flux-engineering pipeline only recovered three transits, and a period of 25.45 days, the 1:3 alias period for Kepler 1811b.

Our flux-only and flux-engineering pipelines detect a new TCE at 554.44 BKJD, shown in Figure 8, which we henceforth refer to as Kepler 1811.02. We detect no other transit-like event within the lightcurve, similar to the new TCE found at 554.44 BKJD (i.e., in shape, duration, and depth). Therefore, this new TCE is a single transit candidate. Similar to Section 4.4, we can place a lower bound on its orbital period, assuming a circular orbit, of 145.9 days based on the transit duration of 10.5 hr. We show the allowed periods for Kepler 1811.02 in Figure 14, accounting for gaps and nongaps in extant Kepler data.

For our MCMC transit fitting, we choose the period 544 days as our initial guess. This corresponds to the shortest period allowed by data gaps for the new TCE at 554.4 BKJD. We run our MCMC, using `batman`, to convergence and find a best-fit radius for Kepler 1811b and 1811.02 of $2.86^{+0.07}_{-0.07}$ and $3.25^{+0.28}_{-0.30} R_{\oplus}$, respectively, assuming a stellar radius of 1.110 $R_{\odot}$. The best-fit parameters from the MCMC transit fit, when assuming a period of 544 days for Kepler 1811.02, are listed in Table 8. The resultant transit model is shown in Figure 8.

The shortest period allowed from data gaps for Kepler 1811.02 is 544 days. The ratio between this period and the

**Table 7**
List of Future Transit Times for Kepler 1897.02 Given the Various Allowed Periods

| $P$ = 342.136 ± 0.004 | $P$ = 559.418 ± 0.142 | $P$ = 589.444 ± 0.089 | $P$ = 665.918 ± 45.555 |
|---|---|---|---|
| 2026-06-23 15:23 ± 1:54 | 2026-03-18 08:49 ± 34:11 | 2026-12-13 14:27 ± 21:39 | 2027-01-04 22:43 ± 9839:60 |
| 2027-05-31 18:38 ± 2:01 | 2027-09-28 18:51 ± 37:35 | 2028-07-25 01:06 ± 23:47 | 2028-10-31 20:45 ± 10933:18 |
| 2028-05-07 21:54 ± 2:07 | 2029-04-10 04:53 ± 40:59 | 2030-03-06 11:45 ± 25:56 | 2030-08-28 18:46 ± 12026:37 |
| 2029-04-15 01:09 ± 2:14 | 2030-10-21 14:55 ± 44:23 | 2031-10-16 22:25 ± 28:05 | 2032-06-24 16:48 ± 13119:56 |
| $P$ = 745.862 ± 0.071 | $P$ = 774.461 ± 0.325 | $P$ = 807.248 ± 0.335 | $P$ = 851.273 ± 1.671 |
| 2026-09-20 15:18 ± 13:46 | 2027-04-08 20:03 ± 62:29 | 2027-11-24 08:16 ± 64:30 | 2026-05-30 05:53 ± 280:58 |
| 2028-10-05 11:59 ± 15:28 | 2029-05-22 07:07 ± 70:16 | 2030-02-08 14:14 ± 72:32 | 2028-09-27 12:26 ± 321:04 |
| 2030-10-21 08:40 ± 17:10 | 2031-07-05 18:11 ± 78:03 | 2032-04-25 20:12 ± 80:34 | 2031-01-26 19:00 ± 361:11 |
| 2032-11-05 05:21 ± 18:52 | 2033-08-18 05:16 ± 85:51 | 2034-07-12 02:10 ± 88:37 | 2033-05-27 01:34 ± 401:18 |
| $P$ = 866.251 ± 0.141 | $P$ = 899.159 ± 1.152 | $P$ = 927.611 ± 0.732 | $P$ = 959.128 ± 0.131 |
| 2026-08-28 02:42 ± 23:55 | 2027-03-13 13:21 ± 193:42 | 2027-08-31 06:32 ± 123:14 | 2028-03-07 08:58 ± 22:10 |
| 2029-01-10 08:43 ± 27:18 | 2029-08-28 17:10 ± 221:21 | 2030-03-15 21:11 ± 140:48 | 2030-10-22 12:03 ± 25:19 |
| 2031-05-26 14:45 ± 30:42 | 2032-02-13 20:58 ± 248:60 | 2032-09-28 11:51 ± 158:23 | 2033-06-07 15:07 ± 28:27 |
| 2033-10-08 20:47 ± 34:05 | 2034-08-01 00:46 ± 276:38 | 2035-04-14 02:31 ± 175:58 | 2036-01-22 18:12 ± 31:36 |
| $P$ = 969.161 ± 0.089 | $P$ = 990.759 ± 0.579 | $P$ = 994.242 ± 0.162 | $P$ = 1019.263 ± 0.233 |
| 2028-05-06 13:42 ± 15:10 | 2028-09-13 03:49 ± 97:24 | 2026-01-13 19:35 ± 23:28 | 2026-05-18 22:02 ± 33:44 |
| 2030-12-31 17:33 ± 17:19 | 2031-05-31 22:01 ± 111:18 | 2028-10-04 01:24 ± 27:21 | 2029-03-03 04:20 ± 39:20 |
| 2033-08-26 21:25 ± 19:27 | 2034-02-15 16:14 ± 125:11 | 2031-06-25 07:12 ± 31:13 | 2031-12-17 10:38 ± 44:56 |
| 2036-04-22 01:17 ± 21:36 | 2036-11-02 10:27 ± 139:05 | 2034-03-15 13:01 ± 35:06 | 2034-10-01 16:56 ± 50:31 |

| $P$ = 1026.436 ± 0.214 | $P$ = 1067.486 ± 0.782 |
|---|---|
| 2026-06-23 18:53 ± 30:59 | 2027-01-15 00:51 ± 112:43 |
| 2029-04-15 05:22 ± 36:07 | 2029-12-17 12:31 ± 131:28 |
| 2032-02-05 15:50 ± 41:15 | 2032-11-19 00:11 ± 150:14 |
| 2034-11-28 02:19 ± 46:23 | 2035-10-22 11:51 ± 168:59 |

**Note.** The future transit times are in YYYY-MM-DD hh:mm ± hh:mm format in GMT. Listed in the header of each square is the allowed period (in days) given a data gap along with the uncertainty in the period, which roughly equates to the width of the data gap. The transit duration of Kepler 1897.02 is 11.47 ± 0.13 hr.

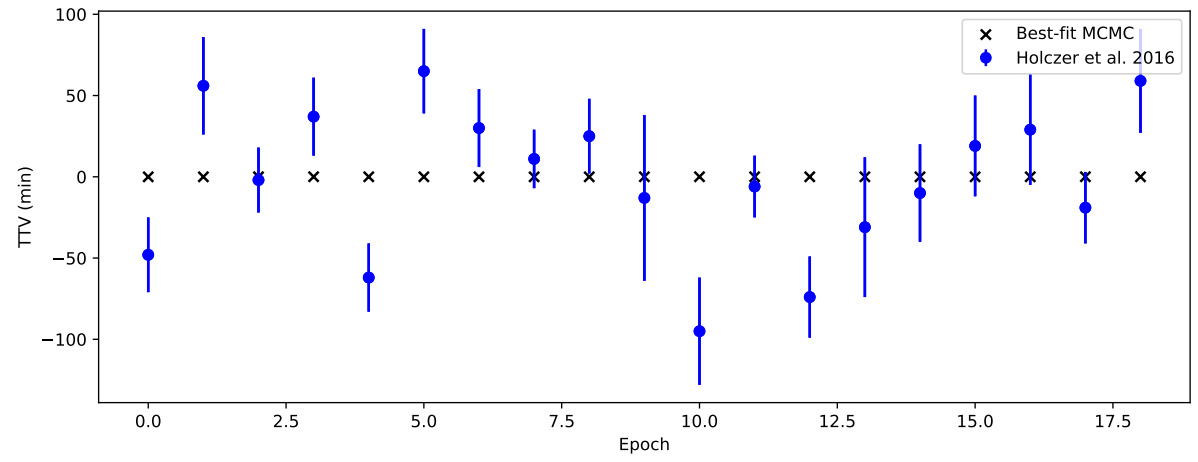

**Figure 13.** The observed TTV signal of Kepler 1811b as reported in T. Holczer et al. (2016) (blue), and the best-fit TTV signal of Kepler 1811b when using a circular model for Kepler 1811.02 produced by `TTVFast` (black). We are unable to reproduce the observed TTV signal of Kepler 1811b with the new candidate Kepler 1811.02 as the perturber with an eccentricity of 0.

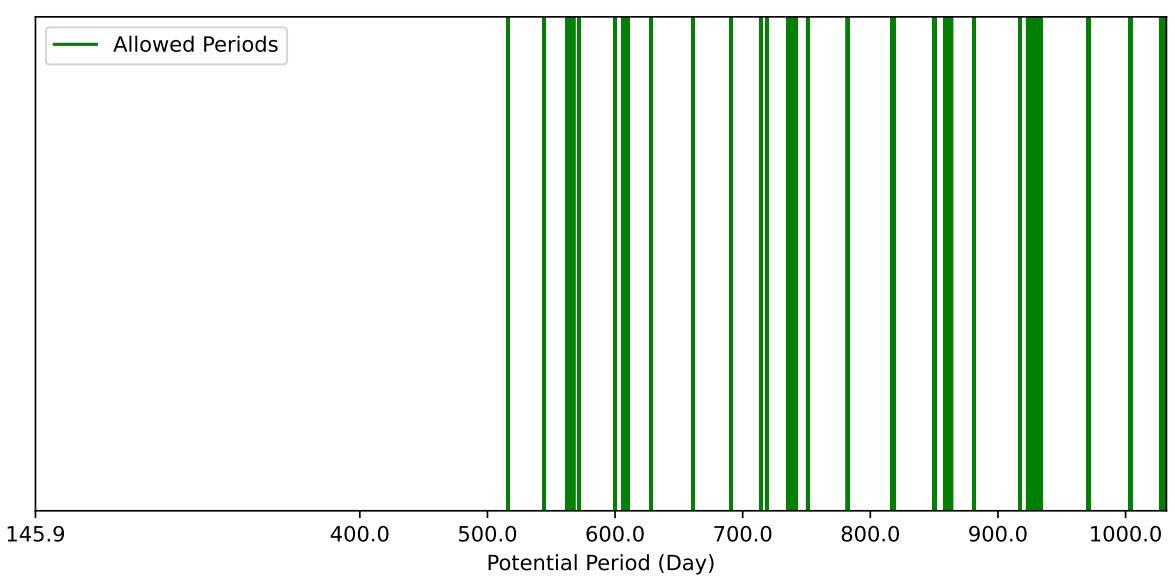

**Figure 14.** The allowed periods with the observed data gaps within Kepler for Kepler 1811.02. The minimum period assuming a circular orbit and an impact parameter of 0 (i.e., the planet transits across the equator of the star) is 308.68 days. The green-highlighted regions are where there are large enough data gaps to hide the other transits of Kepler 1811.02 for that specific period.

**Table 8**
Best-fit Transit Parameters for Kepler 1811, Kepler 1811b, and the New Candidate Kepler 1811.02 with an MCMC and `batman`

| Parameters | Kepler 1811b | Kepler 1811.02 |
|---|---|---|
| $T_0$ (BKJD) | $164.467^{+0.002}_{-0.002}$ | $554.44^{+0.02}_{-0.01}$ |
| Period (Days) | $76.42285^{+0.00001}_{-0.00001}$ | $544.38^{+0.16}_{-0.16}$ [One Potential Period, see Figure 14] |
| Inclination (Deg) | $89.39^{+0.07}_{-0.07}$ | $89.78^{+0.01}_{-0.02}$ |
| $R_p/R_*$ | $0.0237^{+0.0006}_{-0.0006}$ | $0.027^{+0.002}_{-0.002}$ |
| $R_p$ ($R_\oplus$) | $2.86^{+0.07}_{-0.07}$ | $3.25^{+0.28}_{-0.30}$ |
| $\rho_*$ (g cm$^{-3}$) | $0.76^{+0.11}_{-0.11}$ | |
| $gK$ | $0.74^{+0.05}_{-0.05}$ | |
| $hK$ | $0.68^{+0.05}_{-0.05}$ | |

**Note.** We present the best-fit parameters assuming a circular orbit, $e = 0$ and $\omega = \pi/2$. Our best-fit radius for Kepler 1811b is in agreement with the presented values in T. Holczer et al. (2016). To calculate $R_p$, we use the stellar radius value 1.110 $R_\odot$ from T. A. Berger et al. (2026). We report one potential period for Kepler 1811.02 that is allowed by the data gaps within Kepler (see Figure 14 for other potential periods).

known period for Kepler 1811b is roughly 7:1. With no eccentricities, the new candidate cannot cause the perturbations that are observed in Kepler 1811b's transit timings. Figure 13 shows the best-fit MCMC analysis with `TTVFast` when using 544 days as the period for Kepler 1811.02 and masses derived from the power law from J. F. Otegi et al.

**Table 9**
List of Future Transit Times for Kepler 1811.02 Given the Various Allowed Periods

| $P = 515.903 \pm 0.100$ | $P = 544.377 \pm 0.233$ | $P = 563.257 \pm 0.763$ | $P = 567.262 \pm 0.171$ |
|---|---|---|---|
| 2026-01-21 21:01 ± 29:10 | 2026-12-01 02:14 ± 67:32 | 2027-06-26 18:38 ± 220:3 | 2026-01-19 13:37 ± 45:36 |
| 2027-06-21 18:41 ± 31:34 | 2028-05-28 11:17 ± 73:8 | 2029-01-10 00:49 ± 238:21 | 2027-08-09 19:55 ± 49:42 |
| 2028-11-18 16:21 ± 33:57 | 2029-11-23 20:20 ± 78:44 | 2030-07-27 07:00 ± 256:39 | 2029-02-27 02:12 ± 53:49 |
| 2030-04-18 14:01 ± 36:21 | 2031-05-22 05:23 ± 84:19 | 2032-02-10 13:10 ± 274:57 | 2030-09-17 08:30 ± 57:55 |
| $P = 571.768 \pm 0.059$ | $P = 599.956 \pm 0.151$ | $P = 608.293 \pm 1.949$ | $P = 627.889 \pm 0.052$ |
| 2026-03-05 14:57 ± 16:5 | 2026-12-12 12:08 ± 40:10 | 2027-03-05 21:06 ± 514:58 | 2027-09-17 20:10 ± 14:5 |
| 2027-09-28 09:22 ± 17:31 | 2028-08-03 11:05 ± 43:47 | 2028-11-03 04:08 ± 561:44 | 2029-06-06 17:30 ± 15:19 |
| 2029-04-22 03:48 ± 18:57 | 2030-03-26 10:01 ± 47:24 | 2030-07-04 11:11 ± 608:31 | 2031-02-24 14:51 ± 16:34 |
| 2030-11-14 22:14 ± 20:22 | 2031-11-16 08:58 ± 51:1 | 2032-03-03 18:13 ± 655:18 | 2032-11-13 12:12 ± 17:49 |
| $P = 660.841 \pm 0.061$ | $P = 690.889 \pm 0.151$ | $P = 714.470 \pm 0.175$ | $P = 719.109 \pm 0.233$ |
| 2026-10-21 12:23 ± 15:0 | 2027-07-18 22:43 ± 36:33 | 2026-03-03 17:00 ± 38:8 | 2026-04-09 19:37 ± 50:45 |
| 2028-08-12 08:35 ± 16:28 | 2029-06-08 20:03 ± 40:10 | 2028-02-16 04:17 ± 42:20 | 2028-03-28 22:14 ± 56:21 |
| 2030-06-04 04:46 ± 17:55 | 2031-04-30 17:23 ± 43:47 | 2030-01-30 15:34 ± 46:31 | 2030-03-18 00:51 ± 61:57 |
| 2032-03-26 00:57 ± 19:23 | 2033-03-21 14:44 ± 47:24 | 2032-01-15 02:51 ± 50:43 | 2032-03-06 03:27 ± 67:32 |
| $P = 738.440 \pm 3.155$ | $P = 751.088 \pm 0.275$ | $P = 782.250 \pm 0.131$ | $P = 817.916 \pm 0.774$ |
| 2026-09-11 11:12 ± 681:56 | 2026-12-21 15:39 ± 59:43 | 2027-08-27 22:37 ± 28:45 | 2026-03-13 08:33 ± 149:4 |
| 2028-09-18 21:45 ± 757:40 | 2029-01-10 17:46 ± 66:18 | 2029-10-18 04:37 ± 31:54 | 2028-06-08 06:32 ± 167:39 |
| 2030-09-27 08:19 ± 833:23 | 2031-01-31 19:53 ± 72:53 | 2031-12-09 10:36 ± 35:3 | 2030-09-04 04:31 ± 186:14 |
| 2032-10-04 18:53 ± 909:7 | 2033-02-20 22:01 ± 79:29 | 2034-01-29 16:36 ± 38:12 | 2032-11-30 02:30 ± 204:49 |
| $P = 850.302 \pm 0.181$ | $P = 861.131 \pm 2.501$ | $P = 881.318 \pm 0.121$ | $P = 917.127 \pm 0.171$ |
| 2026-10-26 01:22 ± 35:8 | 2027-01-09 20:46 ± 480:39 | 2027-05-31 04:10 ± 23:39 | 2028-02-05 19:59 ± 33:16 |
| 2029-02-22 08:36 ± 39:28 | 2029-05-19 23:55 ± 540:41 | 2029-10-28 11:48 ± 26:33 | 2030-08-10 23:02 ± 37:23 |
| 2031-06-22 15:50 ± 43:49 | 2031-09-28 03:04 ± 600:43 | 2032-03-27 19:26 ± 29:28 | 2033-02-13 02:04 ± 41:29 |
| 2033-10-19 23:04 ± 48:9 | 2034-02-05 06:14 ± 660:44 | 2034-08-26 03:05 ± 32:22 | 2035-08-19 05:07 ± 45:36 |
| $P = 928.499 \pm 5.355$ | $P = 971.051 \pm 0.275$ | $P = 1004.103 \pm 0.325$ | $P = 1029.707 \pm 2.299$ |
| 2028-04-25 10:29 ± 1028:33 | 2026-06-22 05:59 ± 46:32 | 2027-01-06 13:30 ± 54:55 | 2027-06-09 04:34 ± 386:41 |
| 2030-11-09 22:27 ± 1157:4 | 2029-02-17 07:11 ± 53:7 | 2029-10-06 15:58 ± 62:43 | 2030-04-03 21:32 ± 441:52 |
| 2033-05-26 10:25 ± 1285:35 | 2031-10-16 08:24 ± 59:43 | 2032-07-06 18:26 ± 70:30 | 2033-01-27 14:31 ± 497:2 |
| 2035-12-10 22:24 ± 1414:6 | 2034-06-13 09:37 ± 66:18 | 2035-04-06 20:54 ± 78:17 | 2035-11-23 07:30 ± 552:13 |

**Note.** The future transit times are in YYYY-MM-DD hh:mm ± hh:mm format in GMT. Listed in the header of each square is the allowed period (in days) given a data gap along with the uncertainty in the period, which roughly equates to the width of the data gap. The transit duration of Kepler 1811.02 is 10.04 ± 0.13 hr.

(2020). We are unable to produce any magnitude of deviations away from a linear ephemeris for the transit timings of Kepler 1811b.

We calculated the future transit times of Kepler 1811.02. Table 9 lists the predicted future transit times of Kepler 1811.02 for each potential period that is allowed by the visible data gaps within Kepler. We show the next four potential central transit times for each allowed period.

## 5. Discussion

Here we have presented the potential discovery of four new planet candidates within systems containing previously identified planets. Two of the new candidates transit twice within the Kepler dataset, while the other two candidates are single transit events. Kepler 1752.02 and Kepler 199.03 are consistent with periods of $777.78^{+0.01}_{-0.02}$ and $505.495 \pm 0.004$ days, respectively. These orbital periods would make the candidates the 12th- and 67th-longest orbital period candidates for transiting planets to date. Our two single transit candidates, Kepler 1897.02 and Kepler 1811.02, have shortest possible periods of 342 days and 544 days, respectively, that are allowed by the data gaps present in the Kepler dataset. There is a large 90 day data gap within Kepler 1897's lightcurve, which allows for a large potential period range between 621 and 711 days, making these periods the most probable solutions. These four candidates add to the slowly growing population of long-period transiting planets. Currently, of the approximately 4700 transit planet candidates identified by Kepler, 284 have a period greater than 300 days, and only 71 have a period greater than 500 days. In Figure 15, we show the reported planet candidates' radii and periods, along with the four new candidates presented in this paper, with an allowed period range for our single transiting candidates.

### 5.1. In-between Perturbers

Our presented candidates all have longer orbital periods than any other candidates in their host systems. The large separation between the new candidates and their inner companions make them effectively dynamically uncoupled, barring the unlikely case of extremely high eccentricities (and the challenges that come with maintaining long-term stability). However, every inner companion experiences measured TTVs, suggesting the presence of additional companion(s). Unfortunately, none of the systems presented here contain a long enough baseline to accurately measure the period of the TTV signal or determine if there are short-timescale structures within the signal to help extract information on the perturber. In this section, we explore the possible scenario where there is an in-between planet located at the strongest orbital resonances, that is transiting but below our detection threshold, which is causing the observed TTVs of the innermost companions.

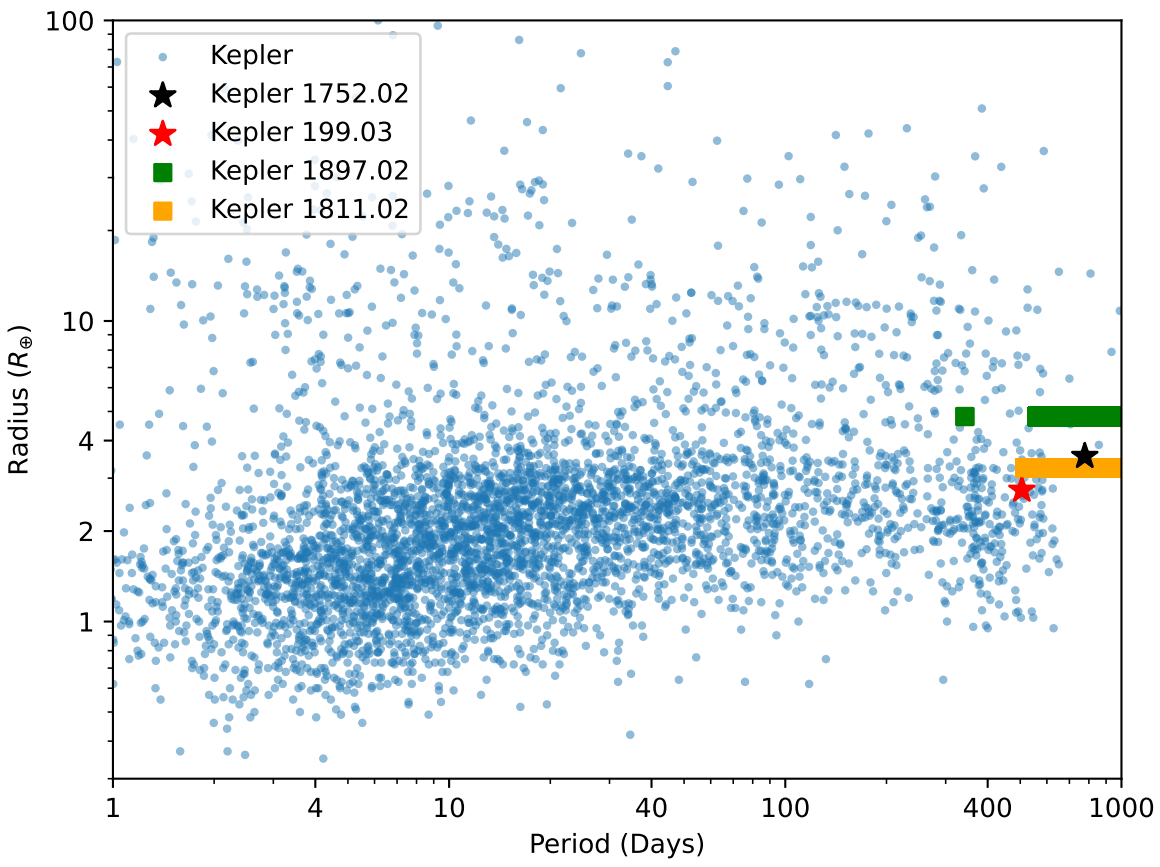

**Figure 15.** Scatter plot of all known planet candidates within Kepler (https://exoplanetarchive.ipac.caltech.edu/docs/counts_detail.html) (blue), along with the four new candidates presented here. Two of the four candidates have two transits within the Kepler dataset (Kepler 1752.02 and Kepler 199.03), and a period is able to be measured (black and red stars, respectively). The other two candidates are single transit candidates (Kepler 1897.02 and Kepler 1811.02). We show the periods that align with data gaps within Kepler as horizontal rectangles. We note that the shaded region of the allowed period for Kepler 1897.02 is noncontinuous.

To determine the size of planets that can be transiting, but went undetected in this and previous studies, we run an injection recovery with astropy's Box Least Squares (BLS) periodogram. We first remove any transit events attributed to the known planets and planet candidates. We then inject at least 5000 planetary signals in the lightcurve to determine the BLS's sensitivity. We utilize the transit modeling package batman to model the injected signal. For the injections, we hold eccentricity to 0, inclination and argument of periastron at 90°, and the limb-darkening parameters to those found in each system in Section 4. We vary the planetary radius in a coarse grid ranging from 0.5–3 $R_{\oplus}$ to determine an initial sensitivity, and then vary the planetary radius in a finer grid about the threshold limit. The injected orbital periods range from 10 days longer than the inner system to 10 days shorter than the outer companion. We grabbed a random time to input into the batman model as our $t_0$, and ensured that the complete transit was visible at this location within the lightcurve (i.e., we do not inject a signal overlapping with data gaps). An injection was reported as a successful recovery if the max peak of the BLS periodogram corresponded to the period within a tolerance of 0.5 day of the injected signal, and contained the random injected $t_0$ within the list of transit times contributing to the BLS peak. We then fit a second-order polynomial representing our detection threshold for each system. The values used for our best-fit polynomial were the first cell within each period column that had a recovery probability of 35% or higher. The parameter space below the line represents a planet that is transiting within the lightcurve data, which would not be detected.

Once we know the radius detection threshold as a function of period, we can associate an upper planetary mass that the potential perturber can have without being detected. To determine if this planetary mass is able to cause the observed TTVs of the inner system, we run two grid searches, where the undetected perturber has a period ratio of either 2:1 or 3:2 with the inner system planet, as these are strong orbital resonances that would maximize the possibility of inducing detectable TTVs in the inner planet. We utilize the *N*-body simulator REBOUND to compute the transit timings for each planet within a system for a given stellar mass, and planetary masses, eccentricities, inclinations, mean anomalies, and argument of periastrons. We use the stellar mass provided by T. A. Berger et al. (2026) for the four stars presented here. For the planetary masses, we use the mass–radius power law for a volatile-rich atmosphere provided by J. F. Otegi et al. (2020). However, the detection threshold for the Kepler 199 system resulted in a small planet, less than 1.0 $R_{\oplus}$, which would not warrant the use of a mass–radius power law corresponding to a volatile-rich atmosphere. Therefore, we use the mass–radius power law for a rocky planet and choose a core-mass fraction of 0.26 for this undetected perturber's mass (L. Zeng et al. 2016, 2019). For the known planets within the systems, we hold all variables as constant. We hold $\sqrt{e}\,\sin(\omega)$ and $\sqrt{e}\,\cos(\omega)$ fixed at a value of 0.2. This provides an eccentricity of 0.04, which we a priori assumed to be enough eccentricity to provide a dynamical interaction, but still a low enough eccentricity to align with previous studies showing multiplanet systems in resonances or containing TTVs remain nearly circular (J.-W. Xie et al. 2016; S. Hadden & Y. Lithwick 2017). The mass of the undetected, transiting planet was selected by choosing a radius below the sensitivity of the BLS threshold that is associated with the 2:1, and then the 3:2 periods.

For each iteration within the grid search, two metrics are taken to determine the accuracy and stability of the proposed solution. Our first metric is how well the proposed simulation's transit times correspond to what is observed. The second is the predicted dynamical stability of the system. To compute the dynamical stability of the system, we utilize the software package SPOCK, which feeds an ML framework statistics from initial *N*-body integrations to predict long-term future stability (D. Tamayo et al. 2020). We use the XGBoost-based feature classifier of SPOCK to provide a probability that the given system is stable over $10^9$ orbits of the innermost planet.

We then sort the grid searches by stability and how well the produced transit times fit what we observe in Kepler. We make a selection cut for the grid searches, where the predicted dynamical stability must be greater than 0.80. This ensures that the remaining parameter space only represents potential stable solutions. We then sort the remaining trials based on their log likelihood values when determining how well they predict the measured transit times. The maximum value corresponds to the best-fit solution that is also predicted to be long-term dynamically stable. The stability requirement is vital, since we are able to produce configurations that reproduce the TTV signal, but have a probability of 0 of being stable.

Once we find the solution with the maximum likelihood, out of all of the stable solutions for a given period, we run an MCMC to fine-tune the parameters. We therefore run two MCMCs for each system, one with the 2:1 period ratio and one with the 3:2 period ratio of the nondetected perturber. We hold tight, uniform priors about the parameters held constant in our grid search, and then wide priors about the three parameters we iterated over in the grid search. The wide priors were the width of the increments in the grid search, i.e, ±5° for the mean anomaly, and ±0.1 for $\sqrt{e}\,\sin(\omega)$ and $\sqrt{e}\,\cos(\omega)$ for the undetected potential perturber. We enforced that the mass of the undetected potential perturber corresponded to a radius below the BLS sensitivity threshold, using the mass–radius relationship from L. Zeng et al. (2016) for the Kepler 199 system, and the mass–radius relationship from J. F. Otegi et al. (2020) for all other systems. Provided a given system configuration, we used REBOUND to produce a list of transit times for all planets within the system. Our log likelihood function computed how well the REBOUND-produced transit timings fit the observed transit times in Kepler for the known planets. We then take the final, best-fit solution based on the MCMC and compute the stability of the configuration as predicted by SPOCK. We do not fold SPOCK into the MCMC for computational time reasons.

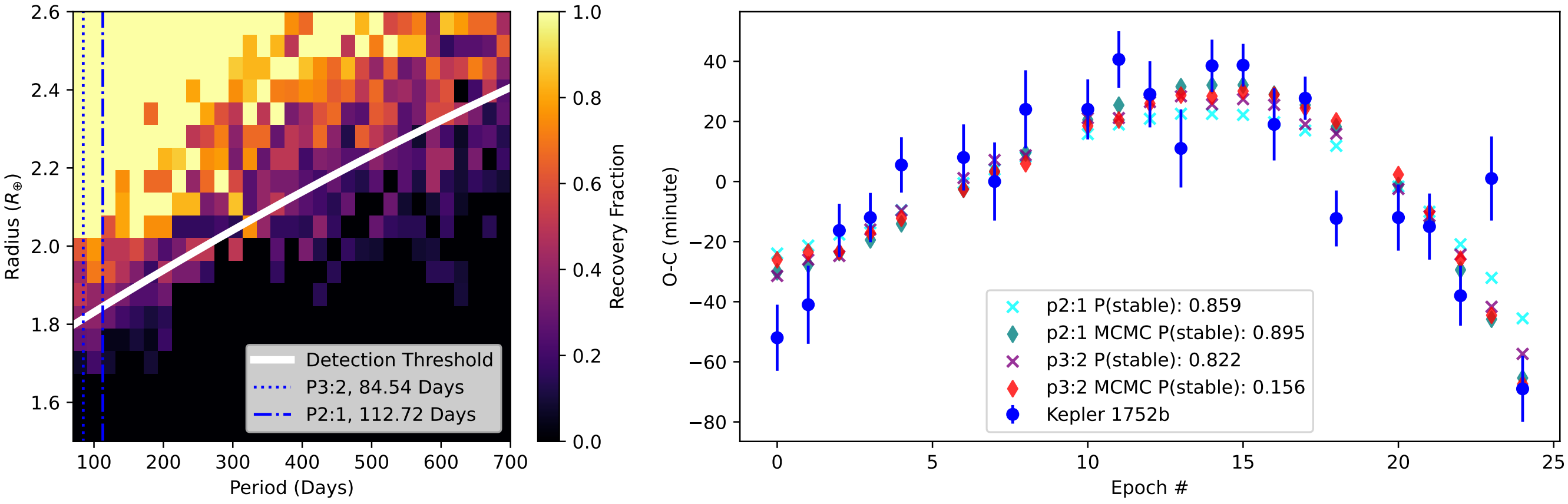

**Figure 16.** Left: the injection recovery of Kepler 1752. A successful recovery was determined by whether the BLS algorithm reported the correct period within 0.5 day and at least one transit from the injection attributed to the signal. The inner companion (Kepler 1752b) has a period of $56.358764^{+0.000008}_{-0.000008}$ days, and the two vertical lines show the 3:2 (dotted) and the 2:1 (dashed–dotted) period for a potential transiting perturber in resonance with the inner planet. The solid white line represents the threshold for which a transiting planet would be detected by the BLS algorithm. The parameter space below this threshold represents the regime where the perturber of the inner companion's TTV is transiting but undetected. Right: the TTV signal of Kepler 1752b as reported in T. Holczer et al. (2016; blue circles); the best-fit stable configurations found from our grid search assuming a 2:1 and 3:2 period ratio for the undetected, transiting perturber (x markers); and the best-fit MCMC-produced solutions about the grid-search locations (diamonds). Also shown is the predicted stability of each configuration, produced by SPOCK. We remove the outlier epoch 5 (see Figure 9) in our analysis.

The results for the Kepler 1752 system are shown in Figure 16. The left plot in Figure 16 shows the injection recovery for Kepler 1752. At the 2:1 and 3:2 period ratios of the inner companion, we are not sensitive to detection for planets up to roughly 1.8 $R_{\oplus}$. We chose a radius of 1.6 $R_{\oplus}$, well below our threshold, to perform our grid search. This equates to a mass of 3.66 $M_{\oplus}$ assuming a volatile-rich atmosphere. The right plot in Figure 16 shows the resultant REBOUND TTV signals compared to the reported signal in T. Holczer et al. (2016). A coplanar, transiting planet is able to be small enough to go undetected from the BLS algorithm, while also causing the observed TTV signal of the inner companion, while maintaining dynamical stability of the system. We note that while the predicted stability of the MCMC for the 2:1 period ratio perturber increases, there is a sharp decrease in predicted stability for the 3:2 period ratio perturber, representing a precarious solution that is able to cause the TTV signal. The final eccentricities for the undetected perturber were 0.75 and 0.53 for the 2:1 and 3:2 period ratios, respectively. The eccentricities for the two known candidates remained around the initial guess of 0.08. Since the resulting REBOUND TTV signal fit the observed signal, we phase-folded the Kepler lightcurve on the produced transit timings of the perturber. We observe no signs of a transiting planet at the 2:1 and 3:2 period ratios. This does not negate these ratios as valid potential solutions but displays the prominent degeneracy in fitting a TTV signal.

The results for the Kepler 199 system are shown in Figure 17. The left plot in Figure 17 shows the injection recovery for Kepler 199. At the 2:1 and 3:2 period ratios of Kepler 199c, we are not sensitive to detection for planets up to roughly 0.9 $R_{\oplus}$. We chose a radius of 0.8 $R_{\oplus}$ to perform our grid search. This equates to a mass of 0.4 $M_{\oplus}$ assuming a rocky planet. The right plots in Figure 17 show the resultant REBOUND TTV signals compared to the reported signal in T. Holczer et al. (2016) for Kepler 199b (top) and Kepler 199c (bottom). A coplanar, transiting planet is not able to be small enough to go undetected from the BLS algorithm, while also causing the observed TTV signal of the inner companion, while maintaining dynamical stability of the system. We do not show the final MCMC results because they had a predicted stability of 0.00, for both the 2:1 and 3:2 period ratios. This is the result of a finely tuned parameter space of dynamical stability within this system. The best-fit solutions from the grid searches produced a stable configuration with eccentricities of 0.5 and 0.4 for the undetected perturber for the 2:1 and 3:2 period ratios, respectively. However, neither ratio is able to fit the observed TTV signal. We also explored a nontransiting configuration, where, for a nontransiting configuration, we can increase the mass/radius of the perturber freely. We were unable to find one that also maintained dynamical stability. Further observations of Kepler 199 are needed to dynamically model this system.

The results for the Kepler 1897 system are shown in Figure 18. The left plot in Figure 18 shows the injection recovery for Kepler 1897. At the 2:1 and 3:2 period ratios of the inner companion, we are not sensitive to detection for planets up to roughly 1.8 $R_{\oplus}$. We chose a radius of 1.6 $R_{\oplus}$, well below our threshold, to perform our grid search. This equates to a mass of 3.66 $M_{\oplus}$ assuming a volatile-rich atmosphere. The right plot in Figure 18 shows the resultant REBOUND TTV signals compared to the reported signal in T. Holczer et al. (2016). There is no clear structure within Kepler 1897b's TTV signal. Nonetheless, a coplanar, transiting planet is able to be small enough to go undetected by the BLS algorithm, while also causing a TTV signal of the inner companion of similar magnitude to the observed signal, while maintaining dynamical stability of the system. We note the decrease in only the 2:1 grid solutions after the MCMC run in predicted stability, where the 3:2 predicted stability increases. The final eccentricities for the undetected perturber, after the MCMC, were 0.07 and 0.54 for the 2:1 and 3:2 period ratios, respectively. The eccentricities for the two known candidates increased slightly to 0.07 for the 2:1 and 3:2 period ratio scenarios.

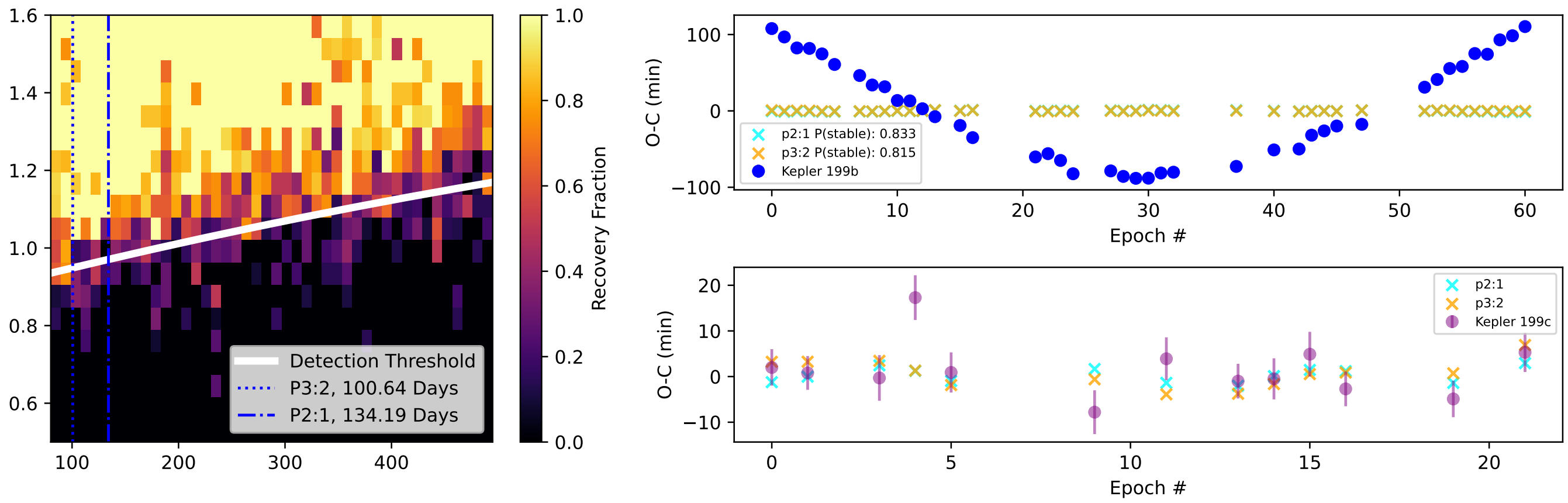

**Figure 17.** Left: the injection recovery of Kepler 199. The outermost, inner companion (Kepler 199c) has a period of $67.093428^{+0.000003}_{-0.000005}$ days, and the two vertical lines show the 3:2 (dotted) and the 2:1 (dashed–dotted) period for a potential transiting perturber. The solid white line represents the threshold for which a transiting planet would be detected by the BLS algorithm. The parameter space below this threshold represents the regime where the perturber of the inner companion's TTV is transiting but undetected. Upper right: the TTV signal of Kepler 199b as reported in T. Holczer et al. (2016; blue circles); the best-fit stable configurations found from our grid search assuming a 2:1 and 3:2 period ratio for the undetected, transiting perturber (*x* markers). Also shown is the predicted stability of each configuration, produced by SPOCK. Lower right: the TTV signal of Kepler 199c as reported in T. Holczer et al. (2016) (blue circles); the best-fit stable configurations found from our grid search assuming a 2:1 and 3:2 period ratio for the undetected, transiting perturber (*x* markers). We do not show the MCMC results, as they had a probability of dynamical stability of 0.00. A transiting planet is unable to be big enough to cause the TTVs, but also remain undetected photometrically.

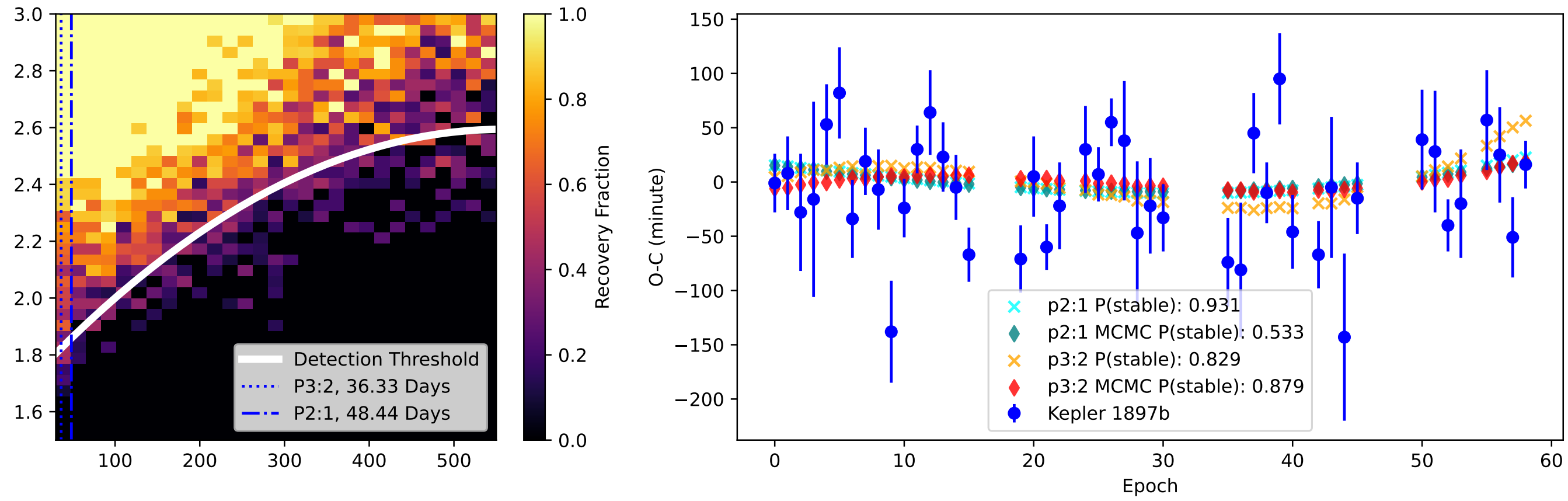

**Figure 18.** Left: the injection recovery of Kepler 1897. A successful recovery was determined by whether the BLS algorithm reported the correct period within 0.5 day and at least one transit from the injection attributed to the signal. The inner companion (Kepler 1897b) has a period of $24.22047^{+0.00001}_{-0.00001}$ days, and the two vertical lines show the 3:2 (dotted) and the 2:1 (dashed–dotted) period for a potential transiting perturber. The solid white line represents the threshold for which a transiting planet would be detected by the BLS algorithm. The parameter space below this threshold represents the regime where the perturber of the inner companion's TTV is transiting, but undetected. Right: the TTV signal of Kepler 1897b as reported in T. Holczer et al. (2016) (blue circles); the best-fit stable configurations found from our grid search assuming a 2:1 and 3:2 period ratio for the undetected, transiting perturber (x markers); and the best-fit MCMC-produced solutions about the grid-search locations (diamonds). Also shown is the predicted stability of each configuration, produced by SPOCK.

The results for the Kepler 1811 system are shown in Figure 19. The left plot in Figure 19 shows the injection recovery for Kepler 1811. At the 2:1 and 3:2 period ratios of the inner companion, we are not sensitive to detection for planets up to roughly 1.5 $R_\oplus$. We chose a radius of 1.3 $R_\oplus$ to perform our grid search. This equates to a mass of 2.63 $M_\oplus$ assuming a volatile-rich atmosphere. The right plot in Figure 19 shows the resultant REBOUND TTV signals compared to the reported signal in T. Holczer et al. (2016). We were unable to fit the TTV signal of Kepler 1811b with a coplanar, transiting planet. Due to the failure of a nondetected, transiting planet causing the TTV signal, we decided to rerun the MCMC, removing the requirement for a transiting configuration and, therefore, also the requirement for the perturber to be below the detection threshold. We let the MCMC explore the complete period range of the injection recovery (90–500 days). The best-fit solution found from the MCMC was not able to completely fit the TTV signal; however, it showed significant improvement over its coplanar counterpart. The best-fit mass for the inclined perturber was 4.75 $M_\oplus$, an eccentricity of 0.27, a period of 143 days, with an inclination of 96°.29. This makes the mutual inclination between the perturber and the inner companion (Kepler 1811b) +6°.81. The predicted stability, from SPOCK, of the mutually inclined solution is 0.707. This is lower than our chosen threshold of 0.80, but still represents a high prediction of stability.

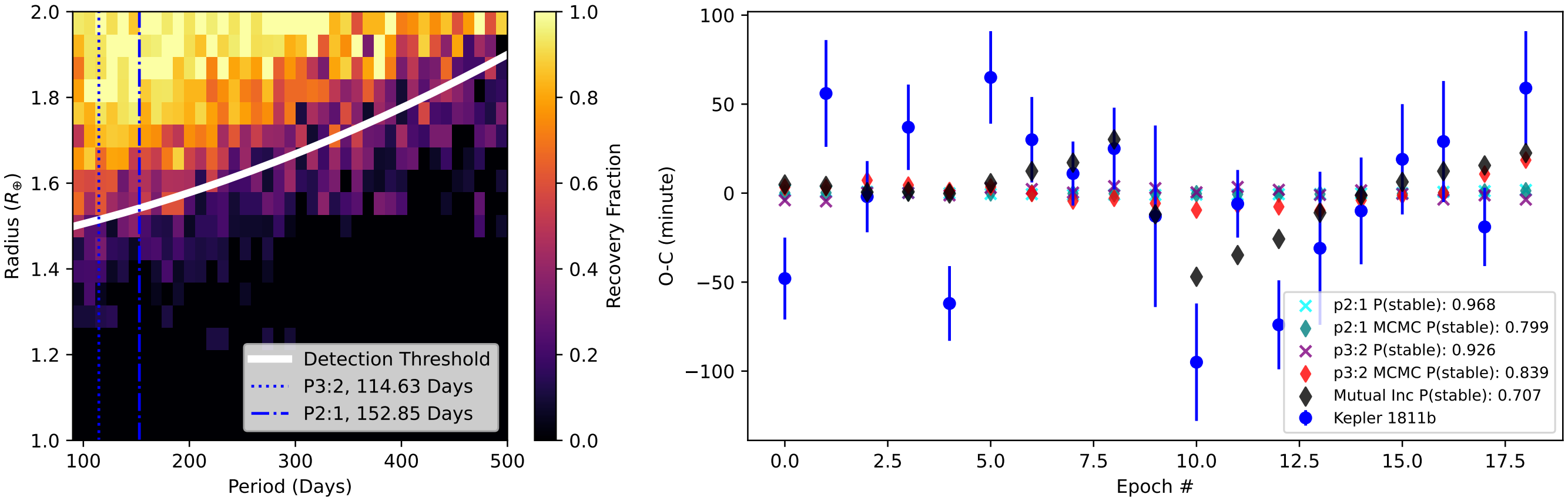

**Figure 19.** Left: the injection recovery of Kepler 1811. A successful recovery was determined by whether the BLS algorithm reported the correct period within 0.5 day and at least one transit from the injection attributed to the signal. The inner companion (Kepler 1811b) has a period of $76.42285^{+0.00001}_{-0.00001}$ days, and the two vertical lines show the 3:2 (dotted) and the 2:1 (dashed–dotted) period for a potential transiting perturber. The solid white line represents the threshold for which a transiting planet would be detected by the BLS algorithm. The parameter space below this threshold represents the regime where the perturber of the inner companion's TTV is transiting, but undetected. Right: the TTV signal of Kepler 1811b as reported in T. Holczer et al. (2016; blue circles); the best-fit stable configurations found from our grid search assuming a 2:1 and 3:2 period ratio for the undetected, transiting perturber (x markers); and the best-fit MCMC-produced solutions about the grid-search locations (cyan and red diamonds). The black diamond is the best-fit MCMC solution when the requirement of coplanar was relaxed and allowed mutual inclination. The mutual inclination solution did not require the perturber to be below the detection threshold. Also shown is the predicted stability of each configuration, produced by SPOCK.

We are able to fit the TTV signals of Kepler 1752b and Kepler 1897b with an in-between perturber, which is coplanar and transiting, but undetected by the BLS algorithm. The solutions were found to be dynamically stable, as predicted by SPOCK. However, there is still a grid of solutions that can produce these observed signals, and we cannot pinpoint an exact configuration for the undetected perturber. This is best seen in the example of Kepler 1752b, where the 2:1 and 3:2 period ratio solutions for the perturber produce roughly the exact same TTV signal. Further observations of Kepler 1752b's transit events may allow for a more precise TTV period signal, and observations of short-timescale TTV structures may help break this degeneracy.

For Kepler 199 and Kepler 1811, we were unable to fit the observed TTV signals with a coplanar, transiting planet that remained small enough to go undetected by the BLS algorithm. When allowing the in-between perturber to become nontransiting and have a mutual inclination with the inner system, we were only able to find a dynamically stable configuration for the Kepler 1897 system. This leaves a wide range of pathways to explore for the Kepler 199 system, which is outside the scope of this paper, in determining the cause of the observed large TTV signal of the innermost planet, Kepler 199b.

### 5.2. Future Missions

Due to the low number of candidate planets at long-orbital periods, pipelines with increased detection efficiency in this regime are able to produce smaller uncertainties when attempting to derive a value for $\eta$-Earth or occurrence rate statistics in general in this region of parameter space. Moreover, populating the long-period regime with new candidates and detailing exoplanet architectures at wide separations may inform future studies to be conducted with Gaia DR4 and Roman. Gaia DR4 (expected no earlier than the end of 2026) and, eventually, DR5 will produce an list of candidate planets detected via astrometry, with the potential of discovering over 100,000 more exoplanets at much larger orbital separation (C. Lammers & J. N. Winn 2026). However, a lack of precise occurrence rates for giant, long-period planets makes these predicted yields highly uncertain. Populating the long-period regime with candidates within Kepler can help constrain these occurrence rates, and help bridge the future occurrence rates from Gaia, which has sensitivity at much larger separations.

Roman is scheduled to launch in late 2026 or early 2027, with one of its science missions being detecting exoplanets via microlensing in the Galactic bulge through the Galactic Bulge Time Domain Survey (GBTDS) (M. T. Penny et al. 2019; S. K. Terry et al. 2026). Roman will also detect exoplanets via the transit method; however, the transiting exoplanet yield will focus on close-in planets (R. F. Wilson et al. 2023). The microlensing and free-floating planet yields focus on large-separation planets (M. T. Penny et al. 2019; S. A. Johnson et al. 2020). Two of the requirements for the GBTDS are to measure the mass function beyond 1 au with masses greater than 1 $M_\oplus$ and less than 30 $M_\oplus$, and an estimate of $\eta$-Earth (S. K. Terry et al. 2026). The lower limit of their requirement is pushing the detection threshold of Kepler, and hence, there is some overlap. Enhanced planet detection techniques within Kepler can help increase the area of overlap for these planet populations. Accurate depictions of the occurrence rates within the Kepler field can be used in the future to draw relations for the occurrence rates of far out planets as a function of radial distance from the Galactic bulge.

Long-period candidates within Kepler have and continue to be presented, such as the ones presented here. However, with the few transits within Kepler, follow-up observations are needed to confirm these planets. Unfortunately, there are significant observational challenges with confirming these candidates with future targeted transit observations. For example, our candidate Kepler 1752.02 has a period of $777.78^{+0.01}_{-0.02}$ days, which means it would take years for a chance to observe even a single transit, with no guarantee of good weather on the observing night in question. In addition to the extremely rare transit events of these new long-period candidates, the transit events themselves for all systems presented here are all more than 10 hr in duration. Therefore, you would only be able to observe the ingress or egress of the event, with the assumption that the telescope is able to observe these dim stars below 1000 ppm precision. These challenges effectively rule out ground-based follow-up for the candidates presented here. However, the future space-based PLATO mission allows for the possible follow-up of these long-period candidates. While the observing strategy for PLATO has not yet been finalized, its primary mission is nominally 2 yr. These 2 yr will be divided evenly between the northern hemisphere and southern hemisphere (V. Nascimbeni et al. 2025; H. Rauer et al. 2025). The Long-duration Observation Phase (LOP) in the northern hemisphere has overlap with the Kepler field and would represent an opportunity for following up on these long-period candidates and potentially confirming them. If the above observing strategy is carried out, this would almost ensure PLATO observing these stars during a transit event of these long-period candidates. Moreover, a possible PLATO extended mission may yield a total of 8.5 yr of total observing time, which could include 4 yr of the LOP within the northern hemisphere, ensuring a follow-up transit of these long-period candidates, not accounting for data gaps and downtime. Additionally, the time separation between Kepler and PLATO would enable more precise measurements of these candidates' orbital periods; yet, there would still be some degeneracies if shorter alias periods fall within data gaps or are not observed during that opportunity of the transit event. However, PLATO is designed to observe bright solar-type stars (with a main target range of 8–11 mag), and the expected noise level for PLATO at the magnitudes for the host stars presented here may obscure the transit events. Figure 22 within H. Rauer et al. (2025) shows the predicted noise level in ppm $h^{-1/2}$ for the PLATO mission. When extrapolating out to the magnitudes of the systems presented here, for a transit event with a duration longer than 10 hr, the candidates presented here are likely to be at the edge of the detection limit of PLATO. Kepler 199 is the brightest star of the four systems presented here, with a Kepler magnitude of 13.595. With a transit depth of 1000 ppm, and five potential transit events within the extended 8.5 yr mission lifetime for PLATO, Kepler 199.03 represents the best possibility for a successful transit recovery within PLATO without being dominated by noise.

Our candidates Kepler 1752.02 and Kepler 199.03 are smaller than their inner companions, yet they have an order-of-magnitude longer periods. All planets within these systems (Neptune to sub-Neptune sizes) are beyond the ice line, suggesting that significant dynamic evolution has shaped these systems since their formation. Determining the composition of these planets would provide insight in their metal enrichment and their atmospheric envelope. Additionally, linking the inner system and outer system planetary architectures may provide clues into possible formation channels as well as provide a more complete census on the types of planetary systems that can form and survive to the present era. The inner and outer solar system have had dynamical interactions and exchange of material in the past, so constraining these potential interactions in other planetary systems would help determine how unique our own solar system may ultimately be. Unfortunately, spectroscopic follow-up is not favorable for these systems, as Kepler 1752b and Kepler 1752.02 have a Transmission Spectroscopy Metric (TSM; E. M.-R. Kempton et al. 2018) of 4.88 and 1.55, respectively, when we adopt masses derived with the mass–radius relationship of J. Chen & D. Kipping (2017). Similarly, Kepler 199 is not favorable for spectroscopic follow-up, as Kepler 199b, 199c, and 199.03 have a TSM of 9.99, 7.68, and 4.54, respectively.

While radial velocity (RV) is the optimal choice to constrain the masses and eccentricities of exoplanets, the systems presented in this paper are relatively high-magnitude stars, and long-period planets, which would require extensive RV data to constrain these parameters and makes these systems unfavorable as follow-up targets. For Kepler 1752.02, with a mass of 12.37 $M_{\oplus}$, from the mass–radius relationship from J. Chen & D. Kipping (2017), an eccentricity of 0 and an inclination of $89\overset{\circ}{.}880$, the semiamplitude of an RV signal is 0.98 m s$^{-1}$. Although RV instruments are able to obtain 1 m s$^{-1}$ precision, the Kepler magnitude of 15.79 for Kepler 1752 and a period of $777.78^{+0.01}_{-0.02}$ for Kepler 1752.02 would render this unfeasible for follow-up observations. For Kepler 199.03, the semiamplitude is 0.65 m s$^{-1}$, which is not currently detectable with current instrumentation. The two inner planets of Kepler 199, Kepler 199b, and 199c have predicted semiamplitudes of 2.27 and 1.79 m s$^{-1}$, respectively, which is currently achievable via the Keck Planet Finder (KPF). However, the 0.65 m s$^{-1}$ precision to detect the new candidate, Kepler 199.03, is still not achievable at Kepler 199's magnitude. We note that future 30 m class telescopes will provide an improved collecting area by a factor of 9 compared to, for example, Keck, allowing a similar SNR to Keck at magnitudes 2.4 mag dimmer. However, these observations are still (at best) at the edge of detection, and a dedicated observing program is unlikely to be allocated. Constraints on the inner, known planets' masses and eccentricities may allow for a more robust TTV analysis and potentially derive a mass for the outer, new companions. Moreover, the new constraints may allow for the discovery of nontransiting/nondetected planets in between the inner and outer planets, revealing information about the mutual inclinations of the systems and a more complete census of these systems.

The four new candidates presented here are all (currently) decoupled with their inner companions, representing stars that contain separate inner and outer planet architectures. The current data is insufficient to probe potential dynamical links in between these planets and planet candidates that causes the observed TTV signals and link the outer planets to their inner companions. Therefore, the systems presented here provide a more complete view of the potential exoplanetary system architectures that exist in our galaxy. Techniques that extend the boundary of the detection threshold in period-space are essential as future, more sensitive instruments come online (e.g., Gaia and Roman).

## 6. Conclusion

In this work we presented the updated version of our single transit detection pipeline, first introduced in M. T. Hansen & J. A. Dittmann (2024). We use a single binary classification network built with a CNN block, which takes in a 128 long-cadence flux segment, concatenated with an FCNN block, which takes the corresponding ancillary engineering files for the segment as input. Our network is able to classify if there is a transit within the given segment. We compare our network with a flux-only network, which was built with a single CNN block that takes in the flux segment followed by an FCNN classification block. We find a marginal difference between the flux-engineering network and the flux-only network in classification accuracy.

We applied the flux-engineering and flux-only networks to all stars with a planetary candidate within the T. Holczer et al. (2016) catalog with a period greater than 6 days. We flag a section of a lightcurve as a TCE if our network predicts a transit with greater than 50% confidence. We perform a series of vetting procedures such as if the TCE had a centroid offset or if the TCE was a systematic. We then visually inspected all remaining TCEs for a clear ingress, depth, and egress. For stars with multiple TCEs, we ensures that the TCEs were similar in shape, depth, and duration. We finalized our TCEs with the introduction of four new potential candidates: Kepler 1752.02, Kepler 199.03, Kepler 1897.02, and Kepler 1811.02. Kepler 1752.02 and Kepler 199.03 contain two transit events within their host's lightcurves. Kepler 1897.02 and Kepler 1811.02 contain only one transit event and are single transit candidates. All new candidates presented here are within systems with previously confirmed planets that exhibit perturbations in their transit timings.

Kepler 1752.02 is consistent with a 777.78 day period, which would make it within the top 20 longest-period exoplanet candidates known to date. Kepler 199.03 has two transits 1010 days apart; however, there is a large data gap in between the two transits, allowing for the 2:1 alias period of 505 days. As Kepler 1897.02 and Kepler 1811.02 are single transits, we cannot deduce a period. However, we can place priors on the period by analyzing where there are data gaps within Kepler that another transit can be located. Figures 12 and 14 show the allowed periods for Kepler 1897.02 and Kepler 1811.02, respectively.

We jointly transit-fit all planets within a single system using an MCMC and batman. We assume an eccentricity of 0 for all planets, as there is currently not enough data to precisely constrain the eccentricity. Kepler 1752.02 has a radius of $3.55^{+0.15}_{-0.15}R_{\oplus}$ with a stellar radius of 0.788 $R_{\odot}$. Kepler 199.03 has a radius of $2.74^{+0.05}_{-0.05}R_{\oplus}$ with a stellar radius of 0.927 $R_{\odot}$. Kepler 1897.02 has a radius of $4.81^{+0.20}_{-0.19}R_{\oplus}$ with a stellar radius of 0.786 $R_{\odot}$. Kepler 1811.02 has a radius of $3.25^{+0.28}_{-0.30}R_{\oplus}$ with a stellar radius of 1.110 $R_{\odot}$.

We are unable to dynamically model any system to accurately match its observed TTV signal. Presently, there is insufficient data to fully constrain the orbital and planetary parameters for the systems presented in this paper. Long-term follow-up studies (e.g., RV and/or ground-based photometry studies) are needed to constrain these systems' parameters, such as their TTV periods, eccentricities, and masses; although, we acknowledge the observational challenge in confirming long-period planets.

## Acknowledgments

M.H. would like to thank Leslie Morales for help with the stellar parameters when modeling transits. The authors would like to thank the anonymous referee for providing comments that significantly improved the quality of the paper. We acknowledge the support of the University of Florida Research Computing, specifically for providing the computational resources and supercomputing infrastructure (HiPerGator) that have contributed to the research results reported in this publication.

## Data Availability

All of the Kepler data used in this paper can be found in MAST (STScI 2016).

*Software:* The authors acknowledge the University of Florida Research Computing for providing computational resources and support that have contributed to the research results reported in this publication (http://www.rc.ufl.edu). astropy (Astropy Collaboration et al. 2022, 2013, 2018), batman L. Kreidberg (2015), EMCEE D. Foreman–Mackey et al. (2013), EXOPLANET D. Foreman–Mackey et al. (2021), lightkurve Lightkurve Collaboration et al. (2018), SciPy P. Virtanen et al. (2020), TTVFast K. M. Deck et al.(2014), vetting C. Hedges (2021), REBOUND H. Rein & S.-F. Liu (2012), SPOCK (D. Tamayo et al. 2020).

## ORCID iDs

Matthew T. Hansen https://orcid.org/0009-0002-7768-620X
Jason A. Dittmann https://orcid.org/0000-0001-7730-2240